\documentclass[sn-basic]{sn-jnl}

\usepackage{graphicx}%
\usepackage{multirow}%
\usepackage{amsmath,amssymb,amsfonts}%
\usepackage{amsthm}%
\usepackage{mathrsfs}%
\usepackage[title]{appendix}%
\usepackage{xcolor}%
\usepackage{textcomp}%
\usepackage{manyfoot}%
\usepackage{booktabs}%
\usepackage{algorithm}%
\usepackage{algorithmicx}%
\usepackage{algpseudocode}%
\usepackage{listings} 
\setcitestyle{aysep={}}
\usepackage{placeins}

\def\ltsima{$\; \buildrel < \over \sim \;$}
\def\simlt{\lower.5ex\hbox{\ltsima}}
\def\gtsima{$\; \buildrel > \over \sim \;$}
\def\simgt{\lower.5ex\hbox{\gtsima}}
\begin{document}
\title{Is the composition of the Solar atmosphere unusual, and if so, why?} 

\subtitle{Possible interpretations}



\author[1,2]{\fnm{Bengt} \sur{Gustafsson}}\email{bengt.gustafsson@physics.uu.se}

\affil[1]{\orgdiv{Department of Physics and Astronomy}, \orgname{Uppsala University}, \orgaddress{\street{Box 515}, \city{Uppsala}, \postcode{SE-75120}, \country{Sweden}}}
\affil[2]{\orgdiv{Nordita}, \orgname{Stockholm University}}


\date{Received: date / Accepted: date}

\abstract{The ongoing discussion about the atomic chemical composition of the Sun is commented on. The main focus in this review is on the deviation of the solar composition from that of most other solar-type stars in that its ratio of volatiles (like the elements C, N, O, S, P and Zn) to the refractories (most metals, like Ba, Ca, Ti, Y, Al, Sc and Zr) tends to be higher in the Sun by 10 to 20\%. What does this tell about the formation and evolution of the Solar System? Scenarios in terms of galactic evolution, formation of the pre-solar nebula, of the evolution of the protoplanetary disk, of the engulfing of planets, and of other processes within the Solar System are considered, as well as the evolution of binary stars with similarly different chemical composition. Finally, implications, if any, on the habitability of the Solar System will be commented on.}

\keywords{Sun: abundances, Stars: abundances, Stars: formation, 
Planetary system: formation, Protoplanetary disks, Planet-star interactions,  Galaxy: evolution}

\maketitle

\newpage
\setcounter{tocdepth}{3} 
\tableofcontents

\newpage

\begin{quote}
\textit{When you have eliminated all which is impossible, then whatever remains, 
however improbable, must be the truth.}\\
								Arthur Conan Doyle:
								The Case-book of Sherlock Holmes
\end{quote}

\section{Introduction}
\label{Intro}
Is the Sun a typical star? In a review written more than two decades ago the question was asked \citep{Gustafsson98} and some answers were given: The Sun is unusually massive, more than 95\% of all stars have smaller masses. It is also single, while more than 70\% of all stars are members of binary or multiple star systems \citep{Duquennoy91, Abt75, Duchene13}. Both these circumstances were tentatively suggested to have \textit{anthropic} ``explanations'', in the sense that the existence of observers in the Solar system would have been less probable with a more massive and short-lived host, or a red dwarf with its habitable zone closer to the active stellar surface, or a planetary system threatened by a more variable gravitation field.  As regards the chemical composition of the Sun, it seemed, however, close to ``normal'', and it had for decades been regarded to be so. In fact, the composition of the Sun and the Solar system was often adopted to represent the ``cosmic chemical composition'', typical at least for the Population I in our Galaxy, although it was early realized, for instance by \citet{Suess64}, that these abundance data “may perhaps only apply to our solar system”.

During recent decades very impressive progress has been made in the analyses of chemical compositions of stellar atmospheres, as summarized by \citet{Nissen18}). A primary reason for this is the construction and use of very large telescopes in the 4 -- 8 meter class and efficient spectrometers with excellent detectors. This has made it possible to collect spectra of hundreds of thousands of stars at enough high spectral resolution ($R \equiv \lambda/ \Delta\lambda$ on the order of $10^5$), great spectral coverage (of several hundred nanometers)  and high signal-to-noise (of at least 100 or more), for efficient measuring of accurate strengths of numerous absorption lines. 

As important are the improvements in theoretical modelling of stellar atmospheres and spectra. Here, important early steps were taken in realistic consideration of different complicated physical phenomena such as the effects of spectral lines (blanketing) on the atmospheric structure, of departures from Local Thermodynamic Equilibrium (non-LTE) in excitation and ionization equilibria among atoms and molecules (see \citealt{Scharmer85}), and of convective energy transport and spatial and thermal inhomogeneities and motions in the atmospheres (see \citealt{Stein98, Freytag02}). This modeling requires accurate atomic and molecular data, including photon absorption cross sections and cross sections for excitations, ionization and dissociation and by collisions, as well as line broadening data, and very considerable improvements have also been provided in these respects \citep{Barklem16}. These improvements in models were enabled by new algorithms for solving complex radiative transfer and hydrodynamics problems, and much more speedy computers. 

The accurate analyses of stellar spectra also require knowledge about fundamental stellar parameters, such as stellar effective temperature, radius and mass (or combinations of these) and overall chemical composition (``metallicity'', ofter represented by the iron abundance). These parameters may partly be estimated from the spectra but additional information is very valuable. In these respects the advances in measuring stellar parallaxes though the Gaia satellite, and in radii and masses through asteroseismological and planet finding satellites (CoRoT, Kepler, TESS) have been quite important.  

All this progress, partly independent and partly concerted in many different fields, has led to improvements in stellar abundances, sometimes lowering the errors by an order of magnitude (see \citealt{Asplund05}, \citealt{Asplund09}, \citealt{Asplund21} and \citealt{Lind24}) to typically less than 25\% (or 0.1 dex in the logarithm). Uncertainties of 0.05 dex or more still remain, however, even for the most important elements that dominate the fraction Z of the mass density of heavy elements, C, N, O, Fe, and Ne, in the Sun. This has been of relevance for the much discussed discrepancy between the solar photospheric abundances, and those inferred from helioseismology data on the depth of the convection zone and the sound speed profile, see \citet{Christensen21} and references therein, \citet{Bergemann14} and \citet{Buldgen24}.

\section{Chemical abundances}
\subsection{The absolute chemical composition of the Sun} 

The standard solar composition according to \citet{Grevesse96}, widely used in models of stellar evolution, contributed a weight fraction of elements heavier than helium, Z = 0.017, which matched the solar seismology data well. However, using improved data and analysis \citet{Asplund05a} found solar abundances of C, N and O significantly lower than those of \citet{Grevesse96}; e.g. the logarithmic abundance (scaled to that of hydrogen = 12) of the most important contributor to Z, oxygen, was lowered from $8.83 \pm 0.06$ to $8.66 \pm 0.05$, and similarly the abundances of the other mass-contributors, C, N, Ne and Fe were reduced. The new photospheric value of Z was now 0.0122. This raised ``the solar abundance problem'': the discrepancy between the solar photospheric abundances and those that seemed needed to explain helioseismology data on the depth of the convection zone and the sound speed profile, see \citet{Christensen21} and references therein, \citet{Bergemann14}, also called ``the solar modeling problem'' (\citealt{Buldgen24}).

In several papers Asplund, Grevesse and collaborators refined their analyses with improved modeling spectra in full 3D, using the hydrodynamic \textit{Stagger} models of Nordlund \& Stein, with better atomic data, careful selection spectral lines and gradually taking NLTE into account (\citealt{Asplund09}, \citealt{Asplund21} and references therein). In the latter paper (AAG) the oxygen abundance was given to 8.69 $\pm 0.04$ and Z = 0.0139. \citet{Amarsi21} also published consistent CNO abundances based on molecular lines. 

This development was accompanied by the work by Caffau, Ludwig, Steffen et al. \citet{Caffau08} worked with the \textit{CO$^5$BOLD} 3D models by Wedemeyer, Freytag et al. They found oxygen abundances of 8.76 $\pm 0.07$ which in later papers (\citealt{Caffau11}, \citealt{Caffau15} and \citealt{Steffen15}) was modified to 8.73 $\pm0.06$ and 8.76 $\pm0.02$, as the [O\,I] 630\,nm  and the O\,I 773\,nm  triplet lines were used, respectively. The group also obtained carbon, silicon and iron abundances about 0.05 dex higher than those of the Asplund \& Grevesse group. 
The mass fraction of heavy elements obtained by \citet{Caffau11} was Z = 0.0153.

Yet another group analysed solar photospheric spectra during recent years to estimate absolute abundances, namely that of Bergemann, Lodders, Magg et al., BLM, (\citealt{Bergemann21}, \citealt{Magg22},  \citealt{Bergemann25}, \citealt{Lodders25}), again using \textit{Stagger} models, or \textit{CO$^5$BOLD} 3D models, and the radiative magnetohydrodynamics simulations of \citealt{Carlsson16}.  The recent logarithmic oxygen abundance from \citet{Bergemann25}, is 8.76 $\pm 0.05$, and again the abundances of C, Mg, Si and Fe are about 0.05 dex higher than those of AAG. The value given for present-day Z is 0.0160 $\pm0.0013$. 

Obviously, some convergence between the results of the groups have occurred, and for most chemical elements the groups agree within the error estimates for the individual elements. There are, however, systematic differences remaining between the groups: for 14 elements with abundances independently determined by \citet{Bergemann25}\,and \citet{Asplund21} the mean difference is $0.055\pm 0.009$ dex and in all cases \citet{Bergemann25} \,give the higher value.  The reason(s) for these differences are not clear. A number of differences in analysis methodology should be considered as explanations for the systematic discrepancies, as well as for the deviation from the estimates from solar seismology: \textit{different observations}: AAG results are partly based on the high-altitude observations by \citet{Delbouille73} of the centre of the solar disk while BLM partly relied on Fourier transform flux spectra obtained at lower altitudes though corrected for water absorption, complemented with data from other facilities. \textit{different choices of spectral lines}: AAG included infrared atomic lines as well as molecular lines, while BLM limited the sample of abundance criteria to optical atomic lines, \textit{differences in atomic data}: like in $gf$ values for O\,I lines or data for the modeling of the blends of the [O\,I] lines, and different $gf$ values for Fe\,II lines,  \textit{differences in model atmospheres and spectrum calculations}: and, in some cases, the use of spatial and temporal averages of 3D models, \textit{different methods in comparing models and observations}: the use of equivalent widths (AAG) and on spectral-line fits (BLM), respectively. A detailed comparison of the problems and virtues of these different approaches must be made before a final judgement is possible and the “solar model problem” possibly be proclaimed to be definitively solved or fully convincingly ascribed to problems in solar seismology, e.g. as a result of errors in opacities or interior mixing. Also, in order to reach highest possible accuracy in the analysis of the solar photosphere, NLTE should be introduced also in the calculations of the model-atmosphere and continuous opacities, e.g. for due consideration of the extra contribution of electrons due to over-ionization, affecting the continuous $H^-$ opacity. Also magneto-hydrodynamic simulations should be used, especially for spectral lines formed in the upper photosphere (\citealt{Bergemann21}). Magnetic fluxes on the order 100 G may have significant effects on the abundance estimates from lines also formed in the deeper atmosphere (\citealt{Fabbian12}, \citealt{Deshmukh22}), but is not clear as yet how much the absolute abundance determinations will be affected by that.

Although it is somewhat worrying that the differences between the different solar analyses appear systematic, the diverging results are not of great significance in the present study which will focus on the strictly differential abundance determination for solar twins as compared with that of the Sun. Yet, the remaining problems of the absolute abundances should be carefully analysed and understood, Among the suggested resolutions of the solar modeling problem some may also have implications on the present topic, since they may reflect chemical gradients in the solar interior, resulting from the accretion history of our star.

\subsection{Differential abundance determinations}

In general, stellar abundance analyses are known to be plagued by several systematic errors, in addition to uncertainties in the modeling of stellar atmospheres and spectra, atomic data etc. Even seemingly more trivial problems in the continuum drawing, effects due to blending spectral lines, telluric lines, etc. may be quite severe (see, e.g. \citealt{Nissen18}, \citealt{Jofre15}, \citealt{Jofre19}) and can make the results worryingly dependent on the selection of spectral criteria and methods of analysis.    
The focus here will be on the much more accurate relative abundance determinations that can be achieved in \emph{differential work}, if similar stars are compared. When the fundamental parameters are very similar between the stars in comparison and identical procedures are used in measuring the spectra, one may expect that most of the systematic errors, in line data (including the effects of blending lines and continuum drawing which are necessary to consider for measuring line strengths), and in modeling of atmospheres and spectra, cancel to a great extent. An exception from this, however, may be the effects of blending telluric lines which may be different depending on stellar radial velocity, stellar altitude and humidity at the observation. To minimize these effects it is important that the selection of lines is scrutinized and that the programme stars are observed at different occasions and different altitudes.

From the discussion of \citet{Nissen18} and references cited therein, as well as from \citet{Amarsi19}, \citet{Matsuno24}, \citet{Amarsi22}, \citet{Lind24}, one can estimate realistic \textit{modelling errors} in differential abundance determinations for the solar twins relative to the Sun. For the O abundance, if determined from the 777 nm lines, the differential correction for 3D-NLTE effects when the analysis is made in 1D-LTE may amount to about 0.02 dex for a star which departs by 100 K from the solar temperature, and less for variations of $log g$ or [Fe/H] within 0.1 dex. For the carbon abundances, as determined from the high-excitation 505.2 and 538.0 lines, the effects are less than 0.01 dex. Iron abundances as determined from Fe II lines at wavelengths longer than 500 nm also show effects smaller than about 0.01 dex. The 3D-NLTE calculations of Mn lines by \citet{Bergemann19} also seem to suggest that the differential corrections should be small for solar twins, although they may be considerable for stars with more different fundamental parameters. Most of the studies for other chemical elements have until now been focussed on the metal-poor stars for which the effect are most prominent. For most elements in the present study, where NLTE calculations in 3D models with varying parameter values in the solar range are still missing, the situation is not quite clear yet. However, in view of the results referred to here for elements which have been studied in 3D-NLTE, and from the more frequent 1D-NLTE results (see \citealt{Lind24}, nlte.mpia.de) it is reasonable to assume that the differential methods have diminished the systematic differential errors in logarithmic abundances down to 0.02 for most elements in solar twins or even 0.01 dex in certain cases.

\subsection{Solar twins}    

The great advances in differential abundance analyses during recent decades have opened up new possibilities to study the structure and evolution of the Galaxy, by means of so-called galactic archeology. Also, detailed abundances present new opportunities to explore the history of our own Solar system, by investigating possible markings of the early history of the system in the solar composition. The present review will deal with one class of such differential analyses, namely the comparison between the Sun and \textit{solar twins}.  

\citet{Cayrel81} coined the term ``solar twins'', a term for ``stars which are practically identical with the Sun''. \citet{Melendez06} devised spectroscopic methods to survey such stars and \citet{Melendez09} compared the solar chemical composition with that of a sample of twins. and found the Sun to be unusually rich in volatile elements, or poor in refractory elements. These results led to considerable discussion. The 11 twins were carefully selected to have fundamental atmospheric parameters close to those of the Sun $(|T_{\rm eff}-T_{\rm eff}({\rm Sun})|< 75$ K, $|\log g - \log g$(Sun)$|< $ 0.10 (cgs), $|[{\rm Fe/H}]|< 0.07$). (Here the usual convention is used that the logarithmic elemental abundance ratios within square brackets are normalized on the solar values.) The stellar spectra which were obtained with the Mike Spectrometer at the Magellan 6.5 telescope at Las Campanas, Chile, had high quality with a resolution R = $\lambda/ \Delta\lambda$ = 65,000 and S/N = 450. The analyses were made on a line-to-line basis strictly differentially relative to the Sun  which was represented by asteroid spectra obtained by the same instrument. This, and the close similarity between the twins and the Sun should make the analyses insensitive to the notorious uncertainties caused by systematic errors in the model atmospheres. Further comments on the techniques and possible errors will be given below, for more details see the review by \citet{Nissen18}. 

\begin{figure}[ht]
\includegraphics[width=\textwidth]{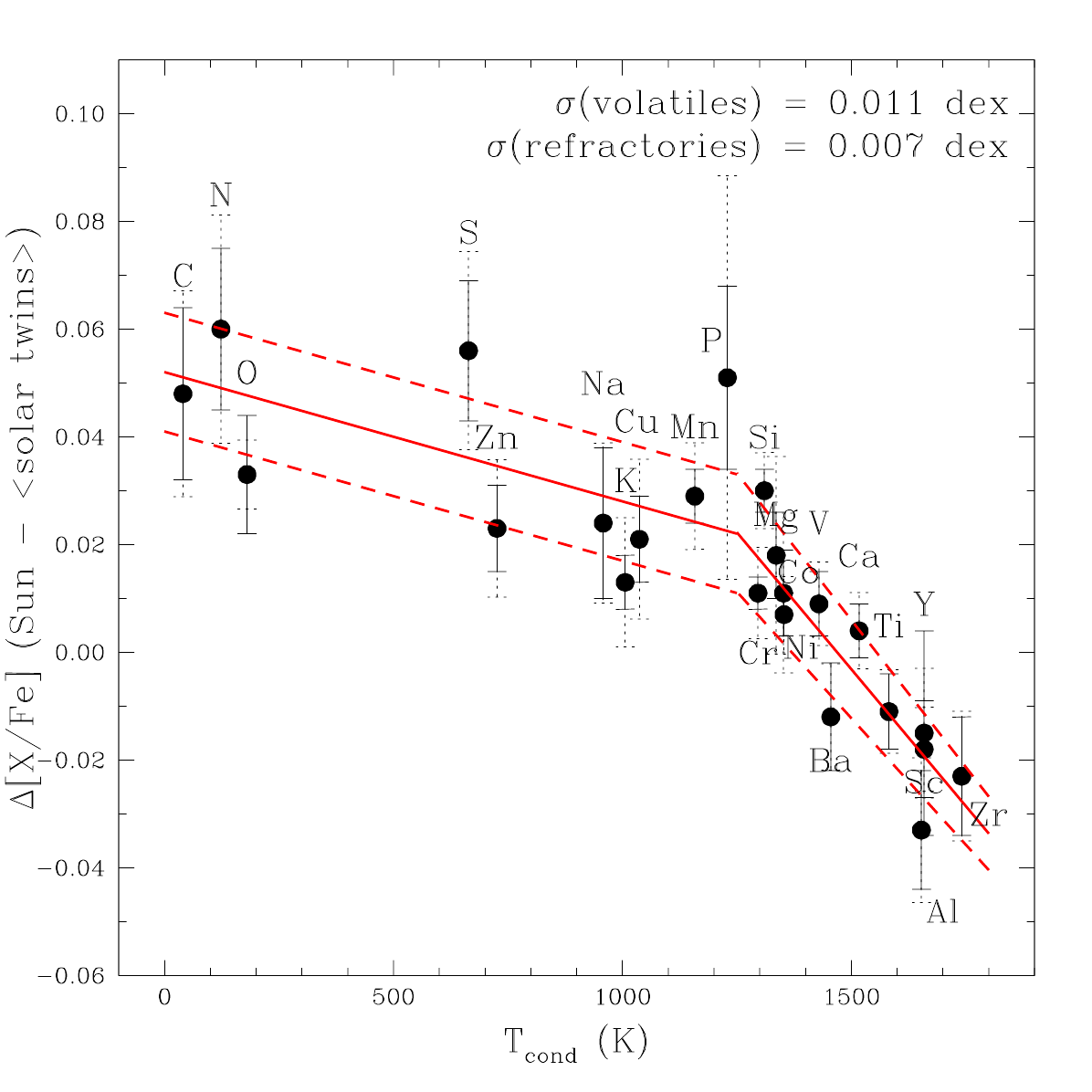}
\caption{The differences between the Solar relative abundance and the mean of the twin abundances, $\Delta_e$ for each element $e$
 versus its condensation temperature $T_c$. A
clear correlation of the abundance differences
with $T_c$  is seen. Regression lines are drawn for elements with low and high $T_c$, respectively. Error bars denote standard deviations in the means, and the spread around the regression lines is indicated in the upper right corner and by dashed lines. Image reproduced with permission from \citet{Melendez09}, copyright by AAS.}
\label{Fig1}
\end{figure}

The resulting abundance differences $\Delta_e \, \equiv $ (Solar $|{\rm X_e/Fe}]$ - mean of 
$[{\rm X_e/Fe}]$ for the twins) for each element $e$
 versus its condensation temperature $T_c$ in a solar composition gas according to \citet{Lodders03}, are plotted in Fig.~\ref{Fig1}. A
clear correlation of the abundance differences
with $T_c$  is seen. The full amplitude in $\Delta_e$ is about
0.08 dex, corresponding to a 20\% variation.
When studying the $\Delta_e$ values for the individual stars it
is found that one star shows an abundance profile similar to the
solar one. These two stars obviously seem to belong to a
minority in this sample of solar-like stars.

\FloatBarrier

The finding by \citet{Melendez09} of the solar-abundance
differences relative to solar twins and similar stars have been
verified by several other studies: \citet{Ramirez10,Ramirez14}, \citet{Adibekyan14}, \citet{Nissen15,Nissen16}, \citet{Bedell18}, \citet{Rampalli24} for
different samples of solar-type stars; see, however
\citet{Cowley22} which will be discussed below in Sect.~\ref{Evolution}. These
different authors typically also find that about 5 --  
15\% of the comparison stars show the solar pattern with high
abundances of the volatiles
(low $T_c$) and low abundances of the extreme refractories
(high $T_c$). The span in these estimates is not astonishing in
view of the differences between the different studies in the
quality of the spectra, differences in comparison stellar samples, in
chemical elements used and in criteria for classifying the ratios
of volaties with respect to refractories as ``normal'' or ``solar
like''. An exception here is the large statistical study by \citet{Nibauer21} 
where about 1700 solar analogues were included with APOGEE spectra in the H band (with a resolution
significantly lower than for other studies cited here, R =22,500),
and the Sun was found to belong to an even larger fraction of refractory-depleted stars. 

It is noteworthy that the cool interstellar gas is also known to
show a systematic depletion of refractories, sometimes by
several orders of magnitude, see, e.g. \citet{Spitzer75}.
This is usually ascribed to condensation of the refractories into
dust in the interstellar medium. One should also note the chemical 
abundance patterns of $\lambda$ Bootis stars, A-type
stars which show considerable depletions of refractories in their
spectra and which make up about 2\% of all A-type stars of
similar temperatures (\citet{Paunzen99}). The thin convection
zone of these stars should make the spectral effects of abundance
changes much more dramatic in their photospheric spectra than
corresponding changes for a solar-type star would show. In
principle, several of the mechanisms to be discussed below
for the solar abundance pattern have also been proposed to
produce $\lambda$ Bootis stars \citep{Venn90,
Waters92, Holweger93, Heiter02, Heiter02a, Kamp02, Jura15}. 

In the following I shall discuss different proposed reasons for the
departing solar abundance pattern, here called the 
\textit{Mel{\'e}ndez effect} (\emph{ME}) as shown in Fig.~\ref{Fig1}. Before entering into such a discussion one must, however, critically examine the possibility that the ME could be an artifact resulting from (1) systematic errors in observations or analysis or
(2) from of an uncontrolled selection bias as regards abundance criteria or stars.

\textit{Systematic errors.} One uncertainty in any differential study of the Sun relative to stars is the enormous difference in energy flux at the Earth from the stars, a difference which makes a direct comparison difficult. A common way to handle this is to represent the solar spectrum by reflected solar light from a solar-system body, like an asteroid, noting that the reflection coefficient of that body does not vary very significantly over short wavelength intervals (\citet{Zwitter07}). Another possible source of error may be that the Sun and bright asteroids show the solar spectrum as observed close to the equatorial plane of the Sun, while the comparison stars may be observed from any angels with respect to their rotation axes. This aspect was explored by \citet{Kiselman11} by spectroscopic observations of the solar disk at different latitudes, and found to be insignificant, only introducing abundance differences of 0.005 dex or less.

Systematic errors in the analyses could result from activity in the atmospheres of the comparison stars to degrees different from that of the Sun. 
Different stellar activity may thus cause 
different magnetic line broadening or heating of the surface layers, which may have spectral
effects, misinterpreted as abundance differences (\citet{Spina20}). However, the $T_c\,$ slopes for
the stars of \citet{Bedell18} are not found to correlate very much with the stellar activity measure
$R’_{HK}(T_{eff})\,$ of \citet{Lorenzo-Oliveira18}, nor do \citet{Spina20} find any evidence in support of the possibility that
chromospheric activity could mimic the \emph{ME}. A heating of the upper solar atmosphere as compared with the twins might cause an effect similar to the \emph{ME}. 
However, such an increased solar activity is not supported by the activity measures of \citet{Lorenzo-Oliveira18}, 
which suggest that Sun is typical for twins of similar age, while the photometric variability measures of 
\citet{Reinhold20} indicate that it is, at least presently, less active than other solar-like stars.

Other systematic errors in the analysis might be due to differences in the fundamental parameters of the stars, which, in spite of the selection of the stars within narrow intervals in $T_{eff}$, log$g$ and [Fe/H], might affect the analyses, for instance due to the only partly known effects of departures from the assumed LTE assumptions and plane-parallel structures of the atmospheres, or the effects of differences in other stellar properties such as rotation speeds or ages. These different aspects will be further commented on below. 

\textit{Systematic biases.} The chemical elements studied here have different atomic masses and ionization energies (of the neutral atom, $\chi_0$), and there is a tendency for increasing masses and for decreasing ionization energies with increasing condensation temperatures and thus with smaller abundance difference $\Delta_e$. There are certainly exceptions from these dependences: for instance Na and K which have low $\chi_0$ and still seem overabundant in the Sun just as Zn and Cu which have relatively great atomic mass, or Al with relatively low mass but small $|\Delta_e|$ and high $T_c$. Also the suitable abundance criteria partly correlate with these circumstances: the availability of relatively weak spectral lines for analysis vary with $\chi_0$\, and $T_c$ and, naturally, abundance. These circumstances could hardly, however, cause the \emph{ME}.   

A more complicated bias may be significant in the selection of comparison stars, which as mentioned above may have had different properties due to different ages, or different chemical profiles due to origins at different formation locations in the Galaxy. Also these circumstances will be discussed below.       

In the following it will be assumed that the \emph{ME}
reflects some processes in the Galaxy, in the early evolution of the solar nebula or the 
solar system, departing from corresponding processes for most twins.
 
Our discussion of will range from effects of galactic chemical
evolution with solar migration in the Galaxy, via dust-gas
separation in star-forming regions and in the pre-solar 
cloud, to processes in the proto-planetary disk, and ingestion of
planets in the stars. The 
latter aspect will also include a discussion about differences
with similar tendencies between
components of binary stars. In the end some speculations
related to the existence of life on
Earth will be commented on.

\section{Galactic processes}	
\subsection{Galactic chemical evolution and solar migration}
\label{Evolution}

The selection of solar twins does certainly not guarantee that the stars and the Sun have common origins in the Galaxy. Even if the ambition were to select stars with similar ages and solar kinematic characteristics, the stars may well have come from quite different galactic regions, possibly with different chemical compositions at different times. 

\begin{figure}[ht]
\centering
\includegraphics[width=0.9\textwidth]{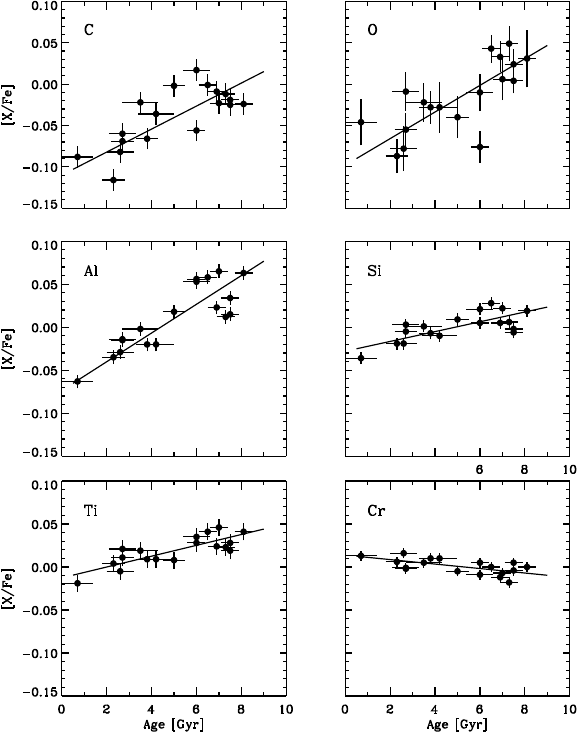}
\caption{Relative abundances for a set of solar-type stars in the Thin galactic disk as a function of isochrone age. Image reproduced with permission from \citet{Nissen15}, copyright by ESO.}
\label{Fig2}
\end{figure} 

\begin{figure}[ht]
\centering
\includegraphics[width=0.8\textwidth]{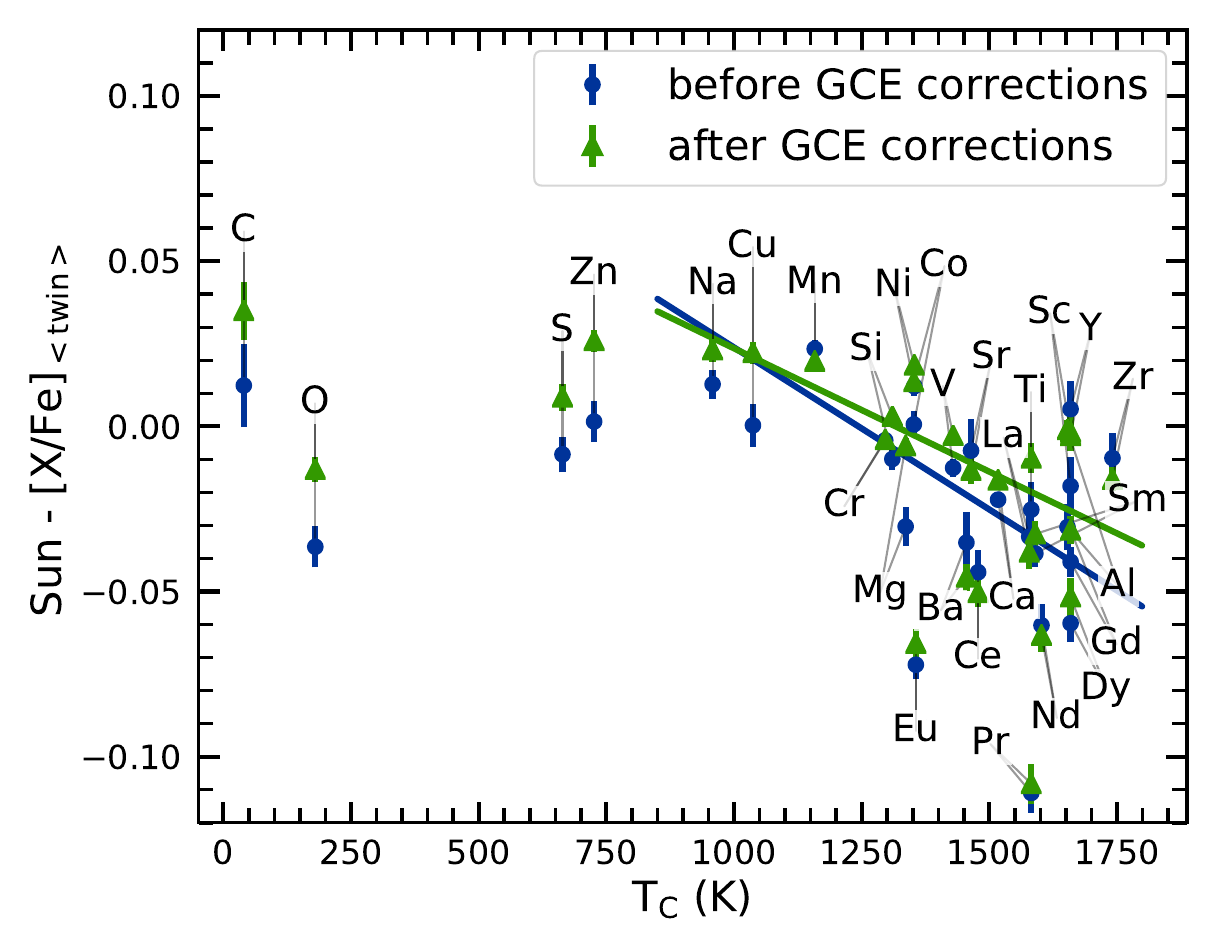}
\caption{The effects on the abundances and the \emph{ME} of correcting the solar-twin abundances for galactic evolution. Image reproduced with permission from \citet{Bedell18}, copyright by AAS.}
\label{Fig4}
\end{figure}

\FloatBarrier
Empirical approaches to take age effects into consideration are based upon estimates of stellar ages from isochrones and distances from the Gaia or Hipparcos satellites, and/or astereoseismic data. The method has for instance been used by \citet{Nissen15, Nissen16}, \citet{Nissen17} and \citet{Bedell18} who plotted abundance ratios vs estimated age for their comparison stars and drew linear relations to correct the abundances to solar age, Fig.~\ref{Fig2} and Fig.~\ref{Fig4}. When extending the sample of stars with ages beyond 5 Gyr, \citet{Nissen20} found two distinct sequences in the abundance-age diagrams so that one linear relation between logarithmic abundance and age was not sufficient for representing the data. 

\begin{figure}[ht]
\includegraphics[width=\textwidth]{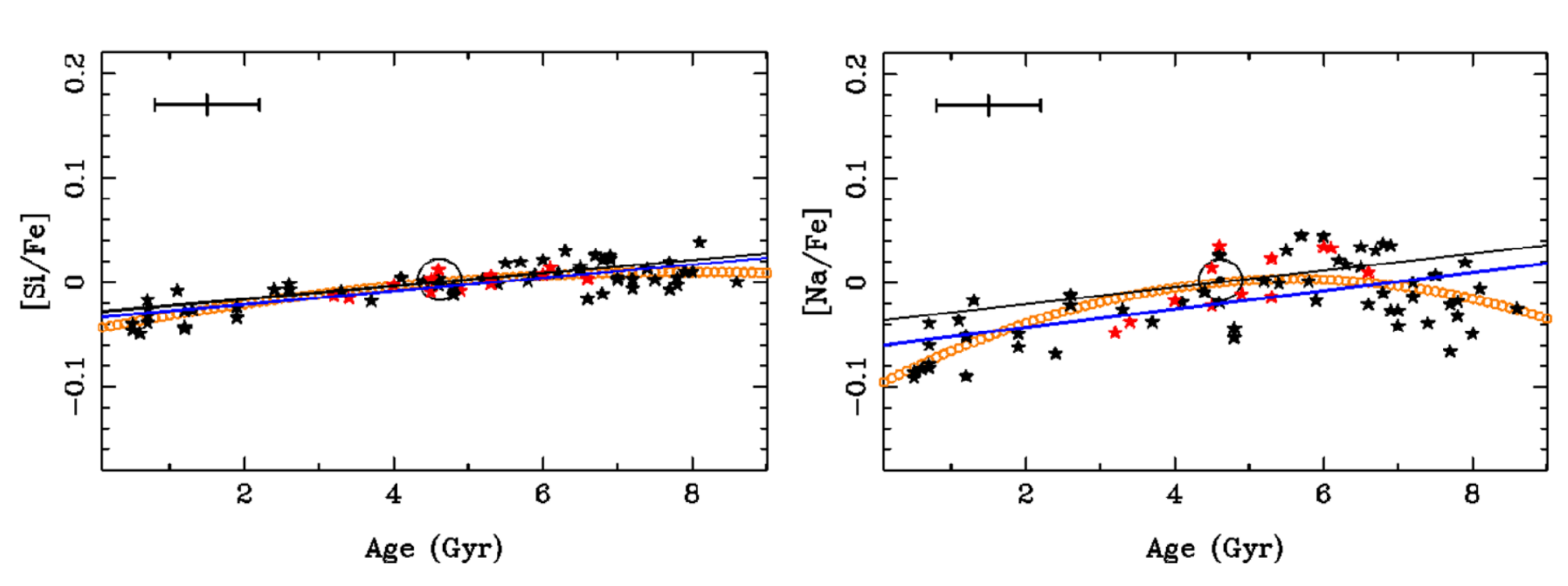}
\caption{Abundance ratios [Si/Fe] and [Na/Fe] plotted for 68 stars from the sample of \citet{Bedell18} relative to stellar ages with linear and quadratic fits. The position of the Sun is indicated. Image reproduced with permission from \citet{Cowley22}, copyright by the author(s).}
\label{Fig3}
\end{figure} 

It should be noted that an observed variation of abundance ratios with stellar age does not necessarily indicate galactic chemical evolution, but may instead reflect a time-dependent inflow of stars or gas of different composition into the solar neighborhood from different galactic regions, or diffusion and sedimentation in the stars or possibly a decline in stellar activity.  
\FloatBarrier

\citet{Cowley22} found that the abundance-age data of \citet{Bedell18} were better represented by second-degree relations and concluded from those that the departure of the Solar abundances from ``twins'' of solar age were non-significant, see Fig.~\ref{Fig3}. It seems, however, that these conclusions are affected by the inclusion in the comparison sample of stars of ages above 6.0 Gyrs. \citet{Tsujimoto20} selected the five stars in the Bedell et al. sample with abundance profiles most similar than that of the Sun, and found that all these stars have ages of about 7 Gyr. They conjectured that these stars, as well as the Sun had migrated from the inner Galaxy as suggested by \citet{Wielen96}, even from the Nuclear Bulge with its presumably different and more rapid history of nucleosynthesis, see e.g. \citet{Baba23}. Significant radial migration should be a common phenomenon in the Galaxy, entailing a considerable fraction of older stars, according to \citet{Lehmann24}. \citet{Adibekyan14} traced a tentative tendency for the $\Delta_e-T_c$ slope to vary with $R_{\rm mean}$, the mean galactocentric radius of stellar orbits, over the interval $6<R_{\rm mean}<9\, \mathrm{kpc}$.

The finding by \citet{Cowley22} that the Solar abundances do not deviate significantly from those of its \textit{contemporary} twins leads to the question why the \emph{ME} shows up at all when linear fits to the abundance-age relations are applied, e.g. by \citet{Bedell18}, see Figure \ref{Fig4}.  It seems that a bending of the true relation at about solar age, as in Figure \ref{Fig3} leads to systematic effects on measures of abundance differences when attempts are made to correct for galactic evolution by means of linear fits. The stronger the curvature of the true relation, the more the Sun would seem to deviate from the typical twins. This hypothesis can be tested by plotting the coefficient of the quadratic fit versus $T_c$ by \citet{Cowley22}, $bb$ in their Table 3. Indeed, a correlation is found, as is seen in Figure \ref{Figny}. Although this would seem to explain the \emph{ME} as an effect of the galactic chemical evolution it also raises the question about the possible physical connection between this evolution and condensation of the gas into solids. If the correlation is just not coincidental, a speculation would be that the dust in the galactic disk is pushed out of the disk by an episode of strong star-formation at some time before the time of the formation of the Solar system. Such an episode might then be connected with the interesting finding by \citet{Nissen20} of two successive sequences of stars in the disk with different abundance ratios lowering the average metallicity significantly about 1 Gyr before the Sun was formed. Although efficient global dust cleansing of gas, presumably by radiation from bright stars, has been traced in early epochs of galaxy formation, see \citet{Jones18}, it is, however, questionable whether the later relatively mild star-formation rate 
according to \citet{Snaith15}, was able to push out dust efficiently at large scales.

\begin{figure}[ht]
\centering
\includegraphics[width=0.8\textwidth]{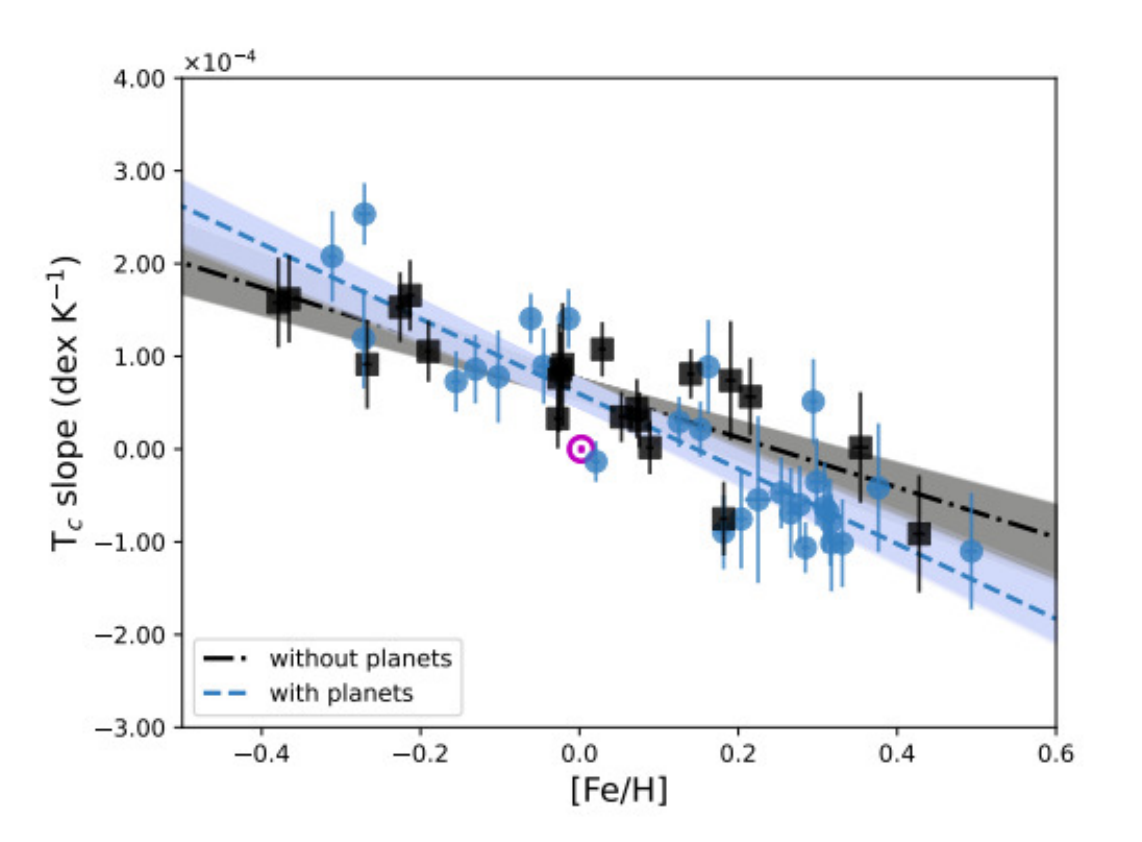}
\caption{The condensation-temperature slopes versus $[Fe/H]$ for 50 F- and G-type stars according to \citet{Carlos25}. Stars with known giant planets are shown in blue, and stars without known giant planets are represented by black squares. The position of the Sun in indicated.  Image reproduced with permission from \citet{Carlos25}, copyright by the author(s).}
\label{Cfig}
\end{figure}  

\FloatBarrier
In a recent study \citet{Carlos25} analysed 50 F- and G-type stars and discovered a clear correlation of the $\Delta_e-T_c$ slope with [Fe/H], see Figure \ref{Cfig}. They ascribe this interesting tendency to galactic chemical-evolution effects. This does not, however, explain the departure of the Sun from its twins: an increase of the solar iron abundance by $\approx 0.15$ dex would have been needed, to get rid of the \emph{ME}. Also, alternative explanations for the tendency may be conjectured, for instance metallicity dependent dust formation in the pre-stellar clouds (see for instance \citealt{Draine07} who find a dust/gas ratio in proportion to the oxygen abundance of a number of galaxies), in combination with the increasing ratio of Fe relative to O with time in the galactic disk and some radiative cleansing by stellar radiation, might cause the correlation of $\Delta_e-T_c$ slope with [Fe/H].

\begin{figure}[ht]
\centering
\includegraphics[width=0.8\textwidth]{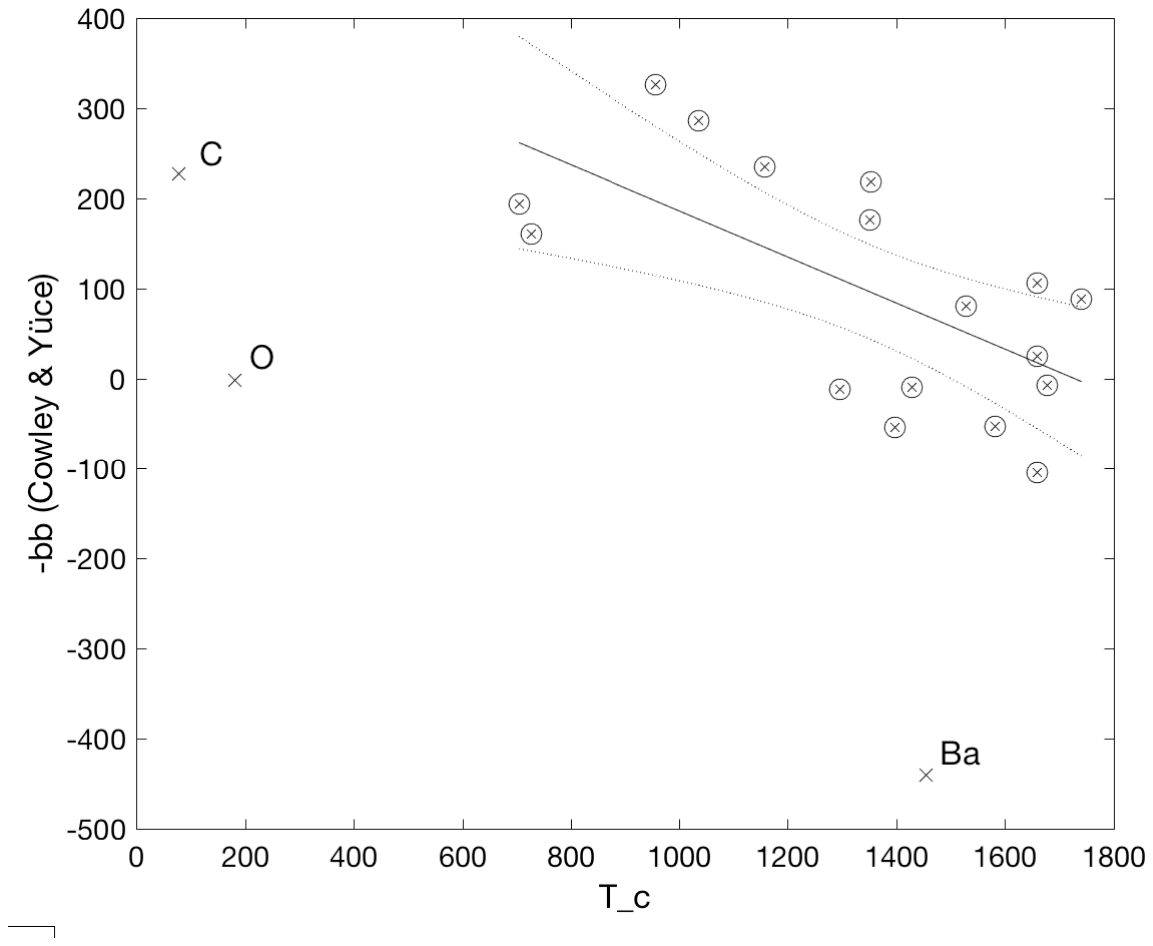}
\caption{The coefficients $bb$ \, of Table 3 in \citet{Cowley22} for different chemical elements plotted versus condensation temperatures from \citet{Lodders03}. The $bb$ values are measures of the curvature of fits of 
[element/Fe] versus age of a set of solar-like stars as adopted from \citet{Bedell18}. The three outliers C, O and Ba are marked by red crosses. A fitted line to the rest of the points, and its $95\,\%$ confidence bounds (dotted lines) are shown.}
\label{Figny}
\end{figure}  

A more theoretical approach to take galactic evolution into consideration in the analysis of \emph{ME} would be to use models of the chemical evolution to correct the observed abundances of the comparison stars to the solar birth age, and to the galactic radius of the assumed birth place of the Sun. Such models of various types and sophistication have been reviewed by \citet{Matteucci21}, see also \citet{Prantzos23}. For recent discussions of the solar neighborhood including migration see e.g. \citet{Kubryk15} and \citet{Johnson23}. Although being dependent on many uncertain parameters such models should be used, together with calculated yields of different elements from supernovae and asymptotic branch stars, to explore whether the particular solar abundances could be fit at all with reasonable assumptions concerning the solar birth place. However, it is uncertain whether a successful match of the observed 
\emph{ME} can be achieved with contemporary yields. \citet{Nomoto13} has compared different sets of yields from SNe. From those it is difficult to explain the \emph{ME} by assuming an atypic population of supernovae that enriched the gas from which the Sun once formed, departing in initial masses from the IMF that led to the composition of the majority of solar-type stars. Especially, problems will show up for
the volatile elements potassium and manganese, which tend to be underproduced, or the refractory 
aluminium which will be predicted to be too abundant. The solar Zn/Al ratio cannot be achieved by postulating any core-collapse SN adding its remnant to the pre-solar cloud, if it contributes elements according to the yields of the models of \citet{Nomoto06} since they all produce too low ratios. However, if Zn is also produced by thermonuclear SN Ia, as has been suggested by \citet{Tsujimoto18}, such a  supernova remnant could possibly match the solar Zn/Al ratio but would then also contribute an excess of Fe.  
\FloatBarrier
\subsection{The local supernova}
\label{Local SN}
The abundances of radioactive atomic nuclei and their daughters in primitive meteorites, suggest the presence within a few pc of a core-collapse supernova in the star-forming region where the Solar system once formed \citep{Cameron73, Looney06}. \citet{Banerjee16} and \citet{Sieverding20} have constructed hydrodynamic models of exploding stars. In the latter work also 3D effects in the inner regions of the model were considered. It was found that this SN must have had a relatively small initial mass to produce the right mix of elements. They suggest a mass of $11.8\,M_{\odot}$, contributing a fraction of about $10^{-4}$ of its debris to the pre-solar cloud. This works well for the radioactive nuclei but adds only small amounts to the general abundance profile. True enough, this addition contains some O, P, S, Na, and K, but also the refractories Al, Si and Sc, which in addition to the small amounts speaks against this single SN as a main explanation for the \emph{ME}.

\citet{Young14} has argued that Wolf--Rayet stars may have produced the radioactive elements traced in the primitive meteorites instead of a supernova. Certainly, such massive stars could have contributed excess C, N and O but hardly S and P, and would probably also overproduce Al \citep{Higgins23}.

\subsection{A dust-cleansed star-forming region}
\label{SF region}

One may ask whether the \emph{ME}, i.e. the depletion of refractories in the solar atmosphere, is an effect of dust formation. Does the observed $T_c$ dependence of the abundances really follow from dust condensation? The degree of condensation reflects the exponential dependence of the equilibrium concentration of n-mers on Gibbs free energy and temperature in the condensation process (see, e.g. \citet{Hasegawa88}, Eq. 9-2-4) and presumably the thermal history of the star-forming cloud.  Empirical indications that the depletion increases with increasing condensation temperature of the element are demonstrated by the great depletion effects in the interstellar gas, which are highly dependent on $T_c$ (see e.g., \citet{Spitzer75}, fig. 5). It is, however, not a simple task to model the \emph{ME} quantitatively for the various scenarios discussed here.

When considering the stellar mass distribution, biased towards stars with small masses, one may conclude that the traces of exploding massive stars close in time from the formation of the Solar system suggest that it formed in a dense stellar environment such as a rich star cluster. One could expect thousands of stars or more within a few pc. Passing stars may have had severe effects on the outer parts of the Solar System and in particular on an early Oort Cloud (see \citealt{Nordlander17}). In general, bright H\,II regions are associated with dust in spiral galaxies \citep{Lynds72}, and star formation near H II regions in the Galaxy show strong dust signatures and varying dust properties \citep{Churchwell09}. In the environment where the Solar System formed, the radiative forces from luminous stars with high masses should have been able to blow away some of the dust and thus cleanse the gas, with the consequence that stars formed in that gas could be depleted in refractories.

We decided to test this hypothesis by determining the abundances in solar-type stars known to be formed in a dense stellar region. We selected the rich open cluster M67, which should early on have contained several tens of thousands of stars \citep{Hurley05} and has approximately a solar age and composition.  It was possible to obtain spectra of satisfactory quality, with the ESO VLT UVES spectrometer, of a few stars at solar brightness and we thus found that the star M67-1194 had a chemical profile more similar to that of the Sun than most of the solar twins known in the Solar neighborhood \citep{Onehag11, Onehag14}. The hypothesis of dust-cleansing by bright stars seemed supported. This particular result was later verified by \citet{Liu16}, who however found also another cluster main-sequence star, M67-1315 with a composition significantly richer in neutron-capture elements than that of 1194.

 In order to further study this dust-cleansing mechanism a simple theoretical model was made of a homogeneous star-forming cloud with a central bright hot star \citep{Gustafsson18a}. The stellar radiation forced the dust grains in the cloud to move radially away from the centre and left gas cleansed from refractories behind. But this \textit{dust front} was soon followed by the \textit{ionization front} and behind that star formation should be much hampered by the high temperature. The cleansed gas in the region between these two fronts could be where refractory-poor stars form, but the total mass available for this was limited due to the thinness of this zone. Possibly a few stars, but certainly not a whole cluster would result. These possibilities were further reduced when taking the turbulence of the gas into consideration: the observed turbulence speeds in molecular clouds are typically orders on magnitude greater than the speed of the dust front in the model, why the cleansed shell in between the fronts must be expected to be swiftly polluted by dusty gas from the cool surrounding shell. 

\subsection{Gas-dust separation in giant molecular and star-forming clouds}
\label{Turbulence}

\begin{figure}[ht]
\centering
 \includegraphics[height=3.8cm]{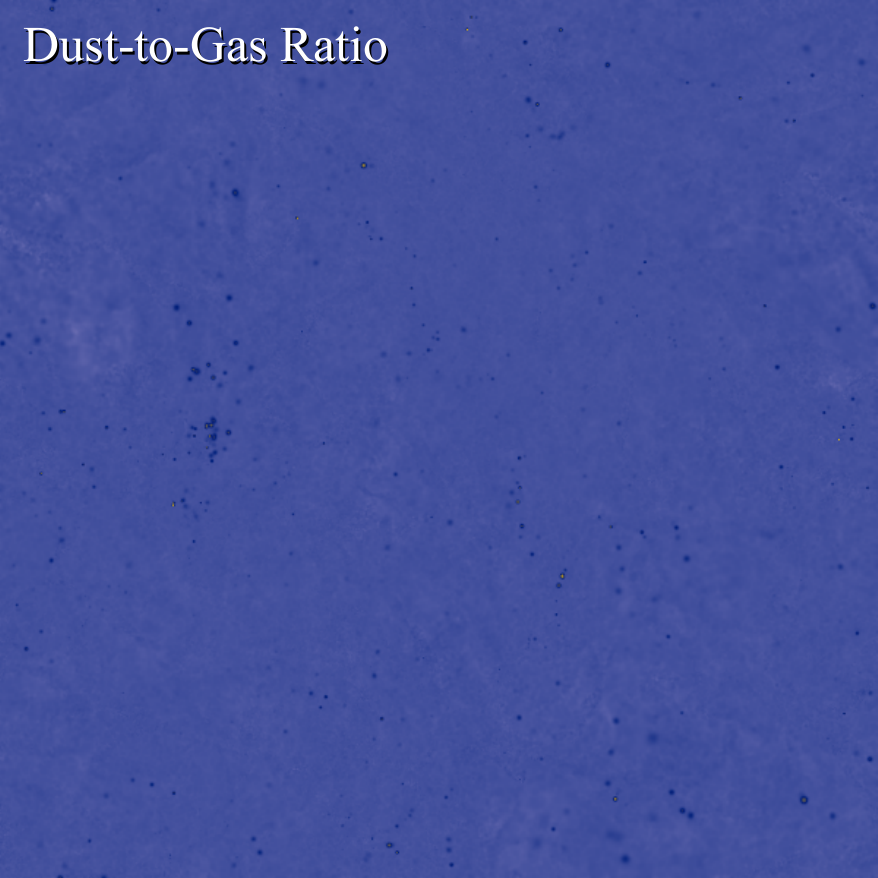} 
 \includegraphics[height=3.8cm]{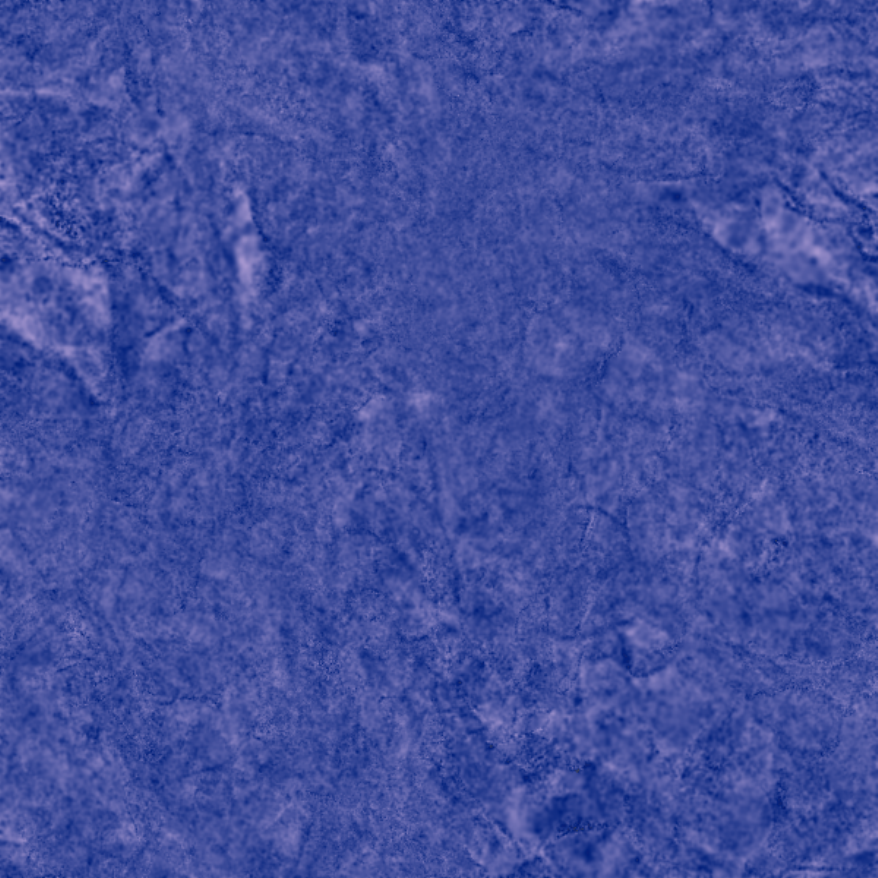} 
 \includegraphics[height=3.8cm]{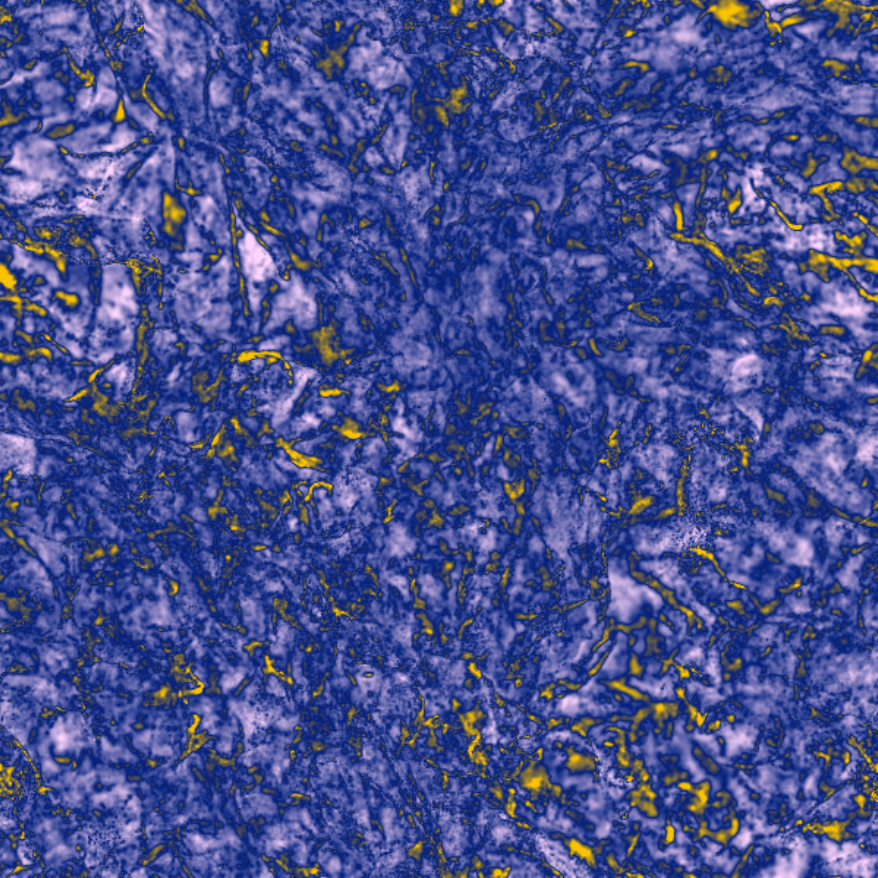} 
 \includegraphics[height=3.8cm]{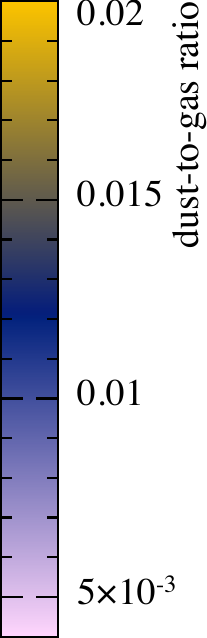}
\caption{Dust-to-gas ratio (mass scale to the right) in the simulations of \citet{Tricco17}, for dust grain diameters of $0.1\mu$ (left), $1\mu$ (middle) and $10\mu$ (right), respectively. The simulation box is 3 pc per side, the turbulence is kept at an rms velocity of Mach 10. Image reproduced with permission from \citet{Tricco17}, copyright by the author(s).}
\label{FigTri}
\end{figure}

Local motions, on relatively small scales (``turbulence''), were long ago proposed to occur in giant molecular clouds by \citet{Zuckerman74}, and have been found to play important roles in the evolution of the clouds. Turbulence, at least when being supersonic, opens up a possibility to produce gas with differences in dust contents, since dust grains of some size could be expected to decouple from the varying motion of the surrounding gas, due to their inertia.  \citet{Hopkins16} thus obtained variations of the dust/gas ratio by orders of magnitude with characteristic mass scales of $1\,M_{\odot}$, in their model calculations. This great effect was, however, demonstrated to be due to numerical artifacts by \citet{Tricco17}, who with more adequate numerical methods found variations in dust/gas ratio of less than 20\% for grain sizes between 0.1 and 1 $\mu$m (Fig.~\ref{FigTri}). However, an increasing dust coagulation will form bigger grains as the star formation proceeds, which might enhance the fractionation process \citep{Bate22}. Later, \citet{Mattsson22} showed that gravity in a cloud could further amplify these effects. It would also be important to include effects of (proto-)stellar radiation fields in such simulations. It cannot be excluded that the observed peculiar composition of the Sun is due to such fractionation phenomena in the pre-solar cloud.

\FloatBarrier
\section{Processes in the protoplanetary disk} 
\subsection{Separation in the disk}
\label{Separation}

\begin{figure}[ht]
\includegraphics[width=\textwidth]{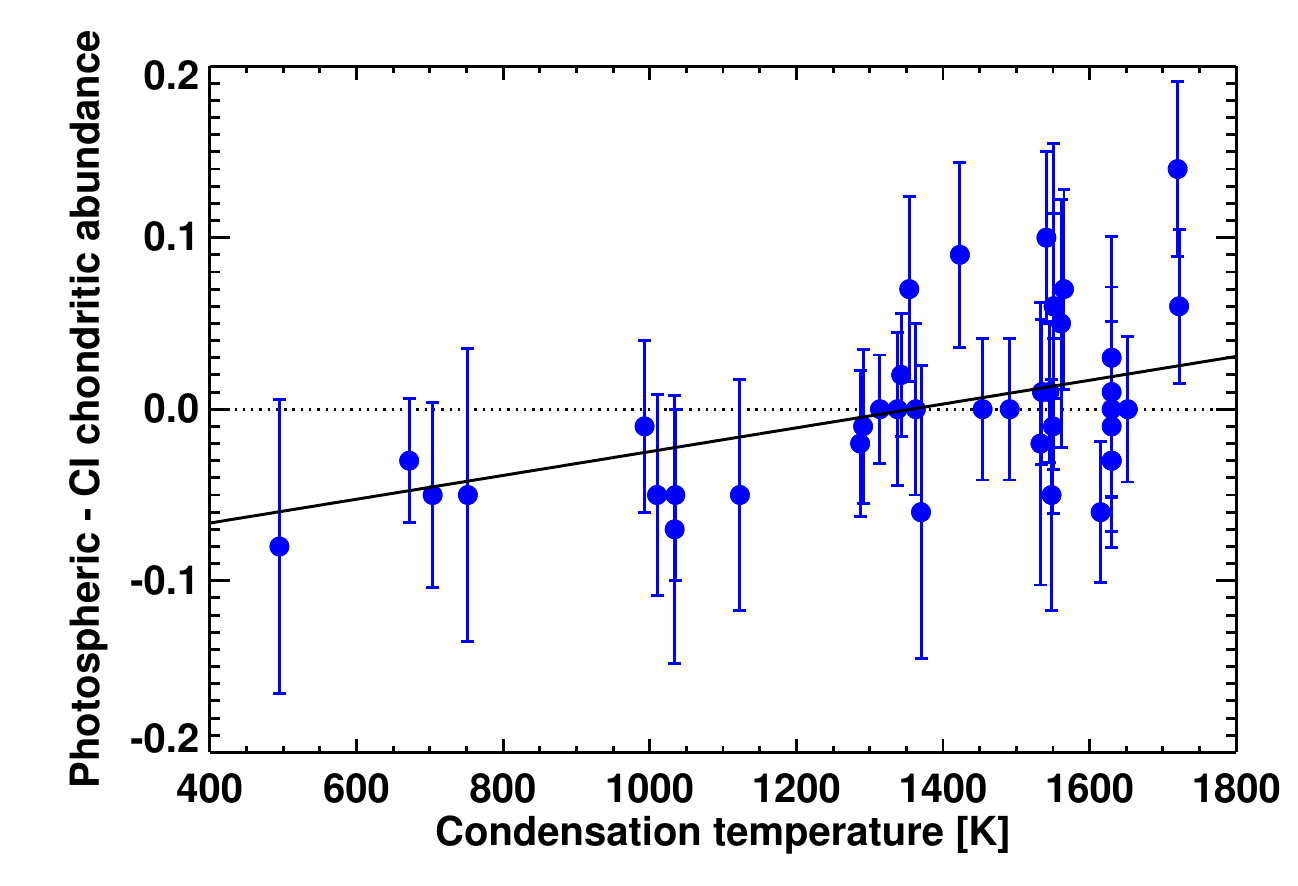}
\caption{The difference between logarithmic photospheric element abundances and corresponding CI chondritic abundances, normalized on silicon, plotted relative to the corresponding condensation temperature for the elements. A regression line is drawn. Image reproduced with permission from \citet{Asplund21}, copyright by the author(s).} 
\label{FigY}
\end{figure}

Different behavior of gas and dust may lead to fractionation of chemical elements in the protoplanetary disk in ways which are yet not fully understood. The carbonaceous chondrites of type I (C I) are stony meteorites with considerable amounts of carbon and combined water with hydrated magnesium-iron silicates, magnesium sulfate and magnetite. The C I meteorites do not show any signs of thermal processing and have since long been expected to reflect the original elemental composition of the pre-solar system cloud \citep{Cameron64}. However, \citet{Gonzalez14}, see also \citet{Asplund21} and \citet{Jurewicz24}, found a systematic deviation (disregarding the extreme volatiles C, N and O and the Noble gases): the volatiles compared with refractories were overrepresented in the C I chondrites as compared with the Sun (see Fig.~\ref{FigY}). The magnitude of this effect seems quantitatively similar to the \emph{ME} but different: the Sun is richer in volatiles with respect to refractories than the solar twins but it tends to be more volatile-poor than the chondrites. This finding was argued against by \citet{Lodders25}, see also \citet{Bergemann25}. If confirmed by further work, the effect could be regarded as an indication that the C I chondrites are not fully representative of the matter in the pre-solar cloud, or that the solar ratio of volatiles versus refractories is lower than in the cloud, although still higher than in the twins. 

Another more striking elemental fractionation becomes obvious when the bulk composition of the Earth is compared with that of the Sun. There are substantial relative-abundance differences (Sun - Earth), systematically increasing beyond one order of magnitude, the lower condensation temperature $T_c$  (see \citealt{Wang19}). 
\FloatBarrier
Similar efficient ``devolatilization'' seems to take place also in different environments such as those in planetary systems around red dwarf stars, presumably with chemical composition not very different from that of the Sun.\citet{Lichtenberg25} have found that the extrasolar rocky planets around M~dwarfs have mean densities rather close to that of the Earth, but note, however, that these data can be fitted with a variety of internal-structure models for the planets.

Different processes may contribute to such fractionation phenomena in the proto-planetary disk: on the microscopic level condensation, evaporation and shattering of dust, and macroscopic radial flows of dust and gas, as well as outflows. After all, the outflows from protoplanetary disks, including magnetized winds may have significant influences on the formation of the star and the disk as well as on the fate of dust with refractory elements \citep{Wang19, Vincovic24}. Also detailed observational studies such as the ALMA-DOT survey of disk-outflow of sources in the Taurus region (\citealt{Codella20} and references cited therein) give interesting information on the distribution of different molecular species in the disks and their outflows. An important question in our context of the \emph{ME} is then what could be the difference between the situation in the early solar disk, compared to those for most solar-like stars.

It should, however, be noted that fractionation in the pre-disk phase or later in the disk would probably imply significant star-to-star inhomogeneities in open clusters. Few such tendencies have been found for the main-sequence stars in open clusters, and in the most well-established case, the Hyades, the limited inhomogeneities found by \citet{Liu16A} do not show a systematic variation with condensation temperature.
  
\subsection{Selective outflows}
\label{Outflows}

\begin{figure}[H]
\includegraphics[width=\textwidth]{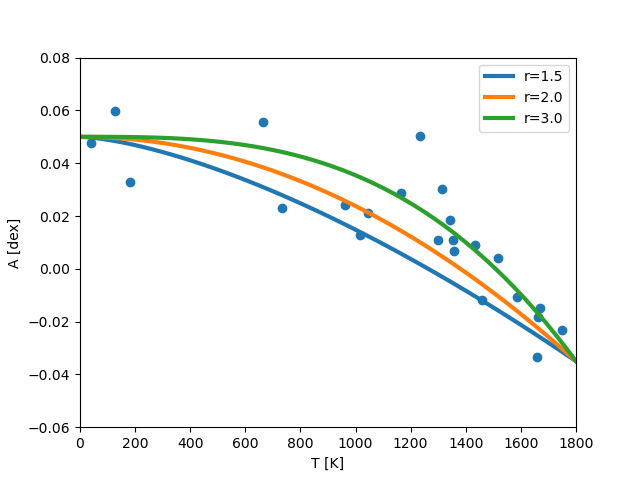}
\caption{The predicted variation with condensation temperature of logarithmic solar abundances relative to twins (here denoted $A$) for the outflow-model of Nordlund (2025) for the protostellar disk. Results for different values of the parameter $r$ of the mass loss rate with distance to the central star are shown. Obviously the curvature of the relation is set by the value of this parameter. Observations from \citet{Melendez09} are indicated by blue dots. A simple normalization has been applied for the amplitude. Image reproduced from \citet{Nordlund25}, copyright by the author(s).} 
\label{FigWnew}
\end{figure}

\citet{Nordlund25} has suggested an interesting model where the abundance effects are the results of the diversity of accretion rates in star formation events, obvious from numerical simulations \citep{Kuffmeier17}, yet all ending with similar final stellar masses. This diversity leads to different conditions in the proto-stellar disks and in selective outflows from them. The temperature gradient in the disk generates sublimation of solids of different composition at different distances from the star, and the resulting gas is contributing to the outflows while much of the remaining solids continue migration inwards to finally enrich the central star. In this schematic model of the Class 0 disk (at the early phase of the formation of the star where it acquires most of its mass) the temperature scales with the radius $a$ as $a^{-3/4}$ which simply follows from the energy balance, and the mass loss per unit radius is estimated to vary as $a^{-r}$ where $r$ is a parameter in the interval 1 to 3. In the model the Sun happened to be less affected as regards its chemical composition than most twins, as a result of its accretion history. For both the Sun and the twins the entire star (not just the convection zone) was enriched in refractories relative to their primordial nebulae, see Fig. \ref{FigWnew}. The model may thus also explain the abundance differences between the carbonaceous chondrites Type I and the Sun discussed above, as well as produce observed differences between the components of some wide binaries to be discussed below in Sect.~\ref{Pairs}.

\FloatBarrier
\subsection{``Self-cleansing''}
\label{Self-cleansing}

A related possibility is that the Sun, and similar stars could have ``cleansed themselves'' from dust (and refractory elements) during their accretion phase. To study this possibility a homogeneous spherical model with dusty gas was set up, with a protosun in the centre \citep{Gustafsson18b}. During which circumstances could the radiation from the star be expected to turn the infalling dust back while the gas would still be accreted? It was found that in order for the dust to decouple from the gas the accretion rate must be small, less than about 1\% of a solar mass per Myr. This mild flow must last for at least 10 Myr in order to produce the reduced refractory markings in the solar spectrum.

A similar though more detailed approach was recently developed by \citet{Soliman24b} in exploring the dust-cleansing in the formation of massive stars ($ > 2.0\, M_{\odot}$).
  
It is worth noting that \citet{Serenelli11} experimented with late accretion flows of magnitudes and longevity similar to those discussed here of solar-evolution models in their attempts to resolve the ``solar modeling problem'': the discrepancy between the solar photospheric abundances from \citet{Asplund09} and those inferred from helioseismolology. Serenelli et al. found that such metal-poor flows may improve the sound-speed profile below the convection zone of the solar model as compared with the standard model and give an excellent agreement with the observed helium abundance, but predict too shallow a convective envelope. Further attempts to model realistic late accretion flows in more detail may be worthwhile.

\section{Effects of planets}
\subsection{Signatures of planetary systems in stellar composition}
\label{Signatures}

Since the first discoveries of extra-solar planetary systems there have been numerous observational attempts to link the existence of planets to the composition of their host stars. Early on, \citet{Gonzalez97} and \citet{Santos01} found that metal-rich solar-type stars tend to be more common among the hosts of giant planets than more metal poor ones. Since then a number of other tendencies have been suggested. One is the suggestion by \citet{Israelian09}, \citet{Gonzalez14} and \citet{DelgadoMena14}, that planet-host stars tend to show lithium depletion in comparison with their non-host counterparts (see also \citealt{Figueira14}). Others, \citet{Baumann10} and \citet{Ramirez12} (for further references see \citealt{Berger18}), could not verify that this effect is due to planets and explained it as results of other stellar properties, such as mixing in the stars due to an unusual high rotation speed of hot-Jupiter hosts \citep{DelgadoMena15}. 

However, planetary engulfments could cause clear Li signatures in the stellar spectra, to a magnitude which is dependent on the planetary mass, the stellar effective temperature (determining the depth of its convection zone) and the time since the engulfment occurred. For a low mass star ($M<0.7\,M_{\odot}$) the signature will vanish within a few hundred Myrs after the engulfment, while for solar mass-stars they may be visible for several Gyrs \citep{Sevilla22}.

High Li abundances found in one components of a binary star as compared with the other component are sometimes accompanied by signs of visible enrichments of refractories; one example is the system HD 240430 \& HD 240429 where \citet{Oh18} found a Li overabundance of 0.5 dex for one of the stars while Fe was enhanced for the same star by 0.2 dex. A detailed comparison of the abundance profiles of the components also disclosed a clear \emph{ME}, see Sect.~\ref{Pairs} below. In their study of 107 binary solar-type stars \citet{Spina21} found that stars with significant abundance differences between the components also had much greater differences in Li abundance as compared with binaries with more homogeneous abundance profiles. Out of the five stars in their sample which showed an excess of Li relative to their companions, all but one also had higher iron abundances. 

\begin{figure}[ht]
\centering
\includegraphics[width=0.49\textwidth]{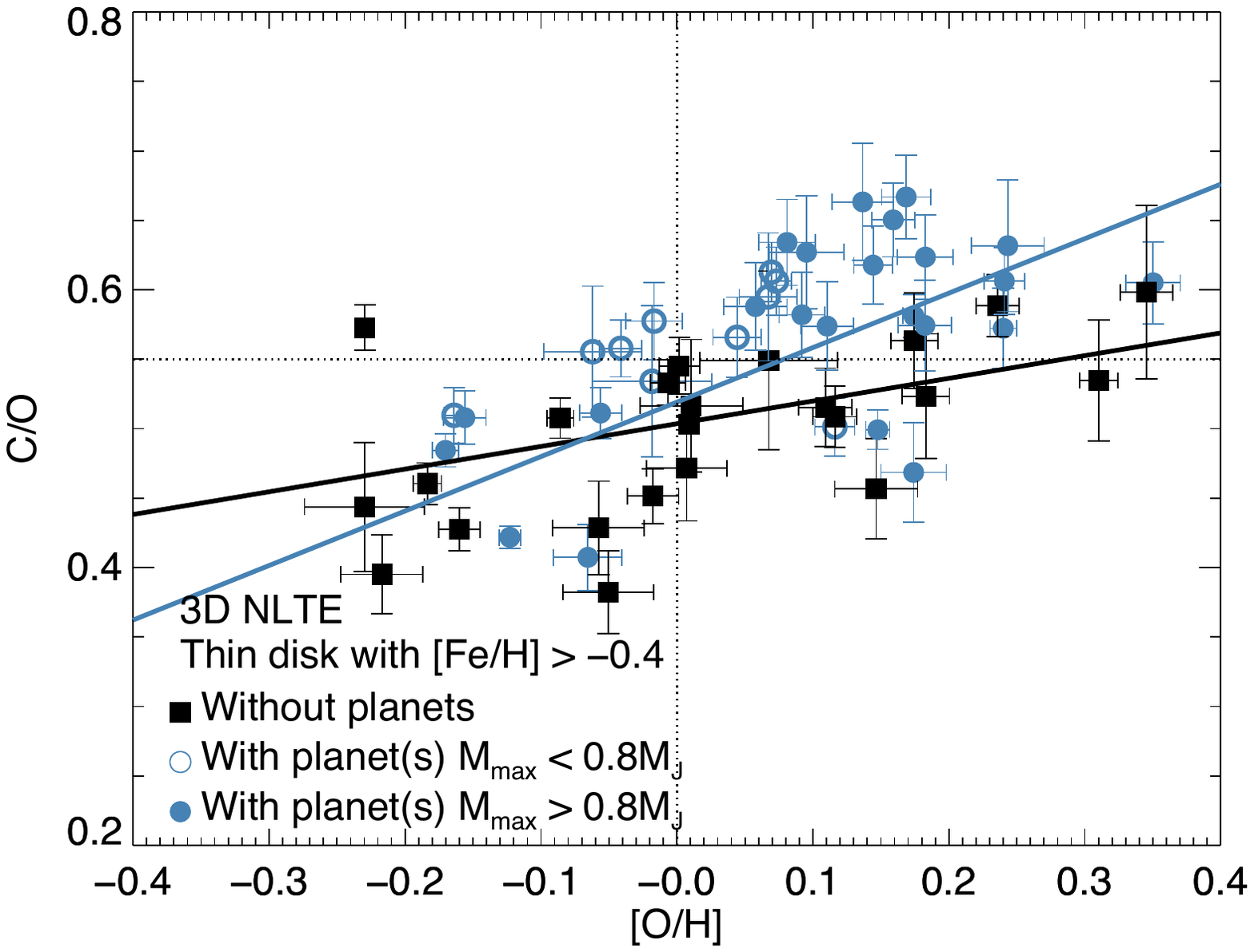}
\includegraphics[width=0.49\textwidth]{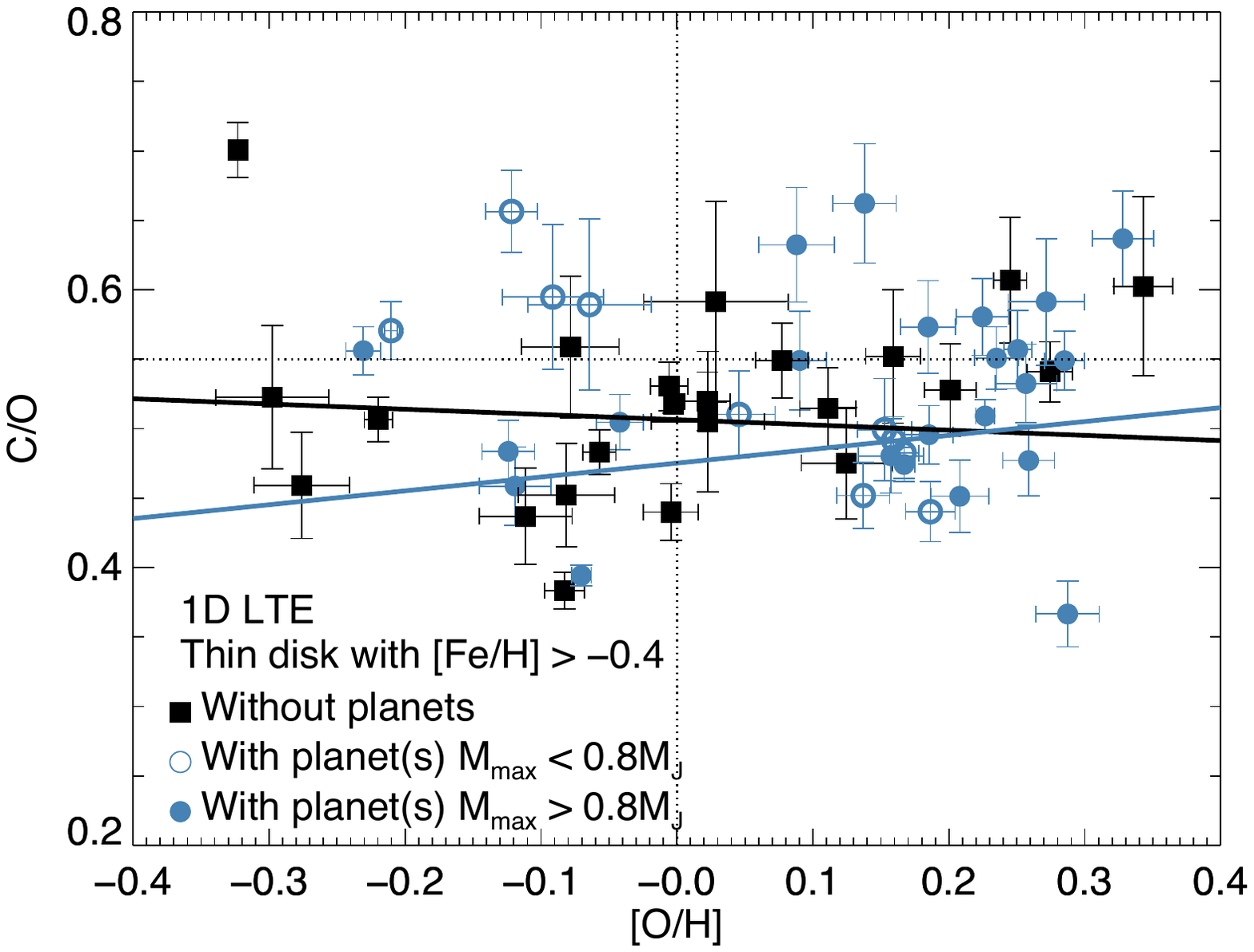}
\caption{C/O radio for thin-disk stars without (black dots) and with planet detections (blue symbols). For the latter stars a subdivision is made into those with a massive known planet and those without (filled and open blue circles, respectively). Results with 3D models and NLTE are shown in the left panel, and with 1D models and LTE in the right panel. Image reproduced from \citet{Amarsi19}, copyright by the author(s).}
\label{FigY4}
\end{figure}

Further findings of abundance signatures indicating the presence of planetary systems, see e.g. the review by \citet{Gonzalez06}, are partly limited by the smallness and bias of the stellar samples used and by the difficulties in making reliable differential analyses of stars with different fundamental parameters. An example of an interesting finding is that of \citet{Amarsi19} that the [C/O] abundance ratios seem to be higher by about 0.1 dex for planetary hosts; in that study 3D models and non-LTE analyses had to be used to disclose the effect by properly taking account of the parameter variations, see Fig. \ref{FigY4}. The interpretation of this finding in terms of planetary system formation is still unclear. 

In the study by \citet{Carlos25} of 50 F- and G-type stars, 29 are known planetary hosts. The authors found no clear correlation of the various [X/Fe]-$T_c\,$ slopes with the presence of giant planets, suggesting that their mere existence does not affect the slope very much, see Figure \ref{Cfig}. This does not exclude the possibility that giants at several AU from the host stars could affect the slopes, as will be discussed below.
 
It must be recalled that our knowledge and statistics concerning extrasolar planetary systems and their host stars is still fragmentary and biased, why conclusions concerning their particular chemical composition are still premature. 

\subsection{The role of gas giants}
\label{Gas giants}

\begin{figure}[H]
\includegraphics[width=\textwidth]{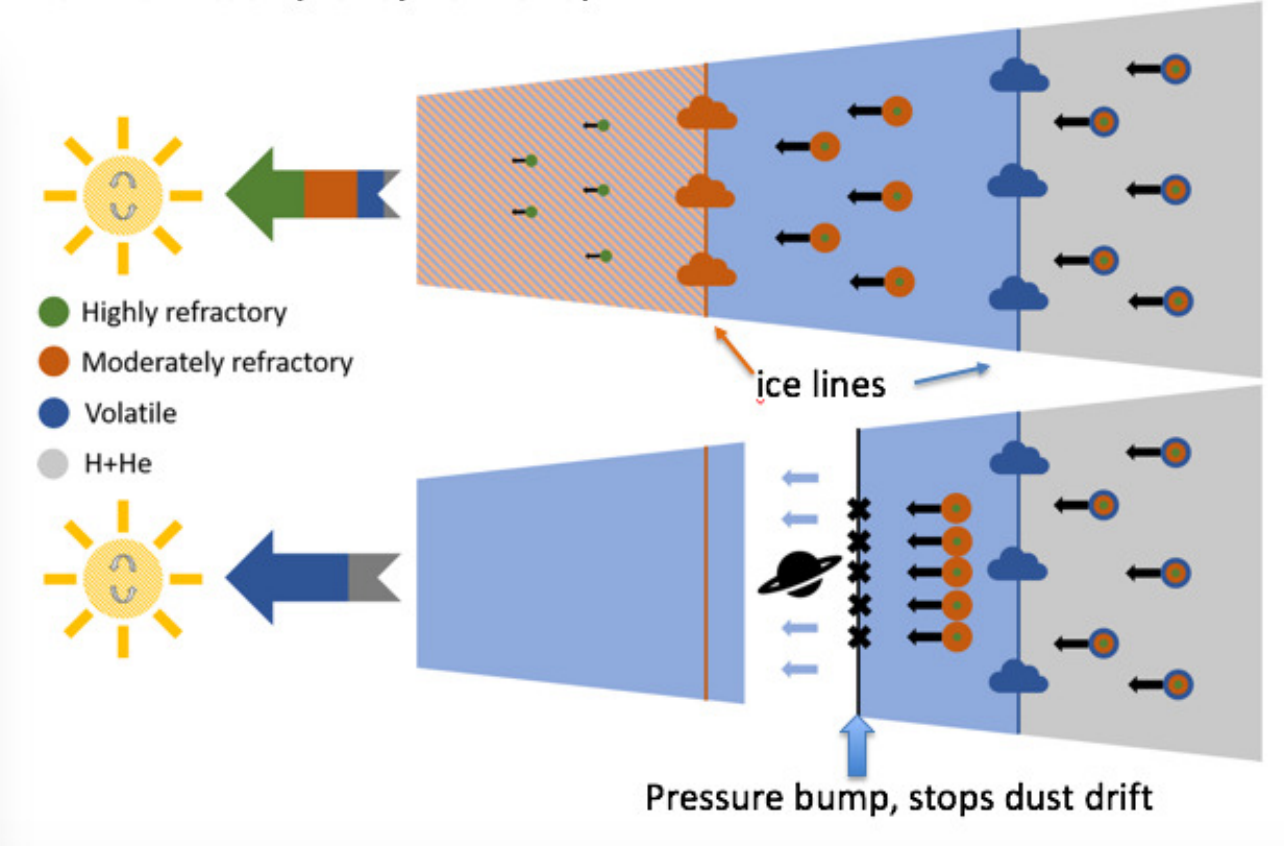}
\caption{The sublimation of different solids at different ”ice lines” in the protoplanetary disk with the unhindered flow towards the Sun (upper panel) and the effects of a giant planet which may stop the drift of solids at the pressure bump (lower panel). Image adapted from \citet{Huhn23}}

\label{FigY1}
\end{figure}

If the giant planet Jupiter formed early, it could open up a ring-shaped gap in the protoplanetary disk, which led to a pressure bump at the outer rim of the gap. Such gaps are observed in about 10--20\% of all disks \citep{Bae23}. If the planet thus formed early enough, so that the central star was still growing by accretion from the disk and the gap was located inside the water ice line in the disk, most of the volatiles would be in gas form while coming to the bump and could stream across the gap and contribute mass to the central star. However, the dust grains or pebbles containing refractories would be halted and trapped by the pressure bump, and as a result the star would become relatively poor in refractory elements. This is the basic idea in the scenario proposed and supported by hydrodynamic models by \citet{Booth20} to explain the \emph{ME}, see Fig.~\ref{FigY1}. In more detailed modelling attempts along these lines with several chemical elements taken into account, \citet{Huhn23} have demonstrated how the results are critically dependent on the time scales involved for the formation of the giant planet (which must be short relative to the life-time of the disk) and for the deep convection zone in the Sun, as well as on the size of the grains. The authors modelled the resulting abundance effects for the Sun and found, with an optimal choice of relevant parameters, solar abundance effects of about half the amount of the \emph{ME}. \citet{Stammler23} have argued, however, that particles trapped at the outer edge of the gap swiftly collide, forming fragments which may then easier pass the gap, why the end results of these processes on the stellar composition are still in doubt.  

In the perspective of \citet{Booth20} and \citet{Huhn23} the special circumstance which made the solar abundances deviate from other solar-type stars would be the existence of a giant planet, Jupiter, in an orbit inside the water ice line, but far enough from the Sun for the dust and pebbles to remain solid. Most of the giant planets around solar-type stars have been found to reside at typical distances of 2--3 AU from their hosts (\citealt{Fernandes19}), probably formed at considerable distances but early migrating inwards due to the friction in their gas disks. In the Solar system a corresponding migration of Jupiter may have been hindered by the near-resonance locking between Saturn and Jupiter.

\FloatBarrier
With this scenario one could expect that stars among the minority with solar abundance profiles, like the M67 star 1194 mentioned above, would have gas giants at distances of several AU. M67-1194 was, however, found to have a hot Jupiter at 0.34 AU by \citet{Brucalassi17}. \citet{Rampalli24} did not find any tendency among their 50 stars with identified close-in giant planets to depart in chemical composition from those of their stars with no known planets. Also, several of the binary components to be discussed below, poor in refractories with respect to their companions, turn out to have hot Jupiters in close orbits. However, these facts are yet not conclusive arguments against the scenario with distant giants trapping pebbles in the disk. One may for instance envisage that the tidal effects by cluster members passing by M67-1194, or by the refractory-rich component of the binaries, have linked the more distant gas giants closer to their hosts long after the evaporation of the protoplanetary disks.

\subsection{Rocky planets as deposits of refractories and indicators of habitability}
\label{Rocky planets}

Early on \citep{Gustafsson10, Chambers10}, it was realized that the amount of refractories missing in the Solar atmosphere could be matched by the amount locked up in the planets. For instance, if the planets Mercury, Venus, Earth and Mars would be mixed into the present solar convection zone the photospheric spectrum would agree much more with the spectra of the twins. The conclusion could then be that the \emph{ME} reflected that the Solar system formed and retained its rocky planets instead of dumping them, or the corresponding material, onto the star. This also led to the suggestion that abundance profiles among stars similar to the solar one could be explored as indicators of presence of rocky planets able to host life \citep{Melendez10}.

Soon it was, however, realized that this idea that the refractories missing in the Sun were deposited in the rocky planets around it had considerable caveats. In particular the time scale of the protoplanetary disks was estimated from observations of pre-main-squence objects to be less than 10 Myr \citep{Alexander14} while it would take more than 25 Myr for the deep hydrogen convection zone of the early Sun to get thin enough to only contain a few percent of the solar mass as it does today. An accretion before 10 Myr on the Sun of the remains of the protoplanetary disk cleansed from refractories by planetary formation would then hardly lead to any visible effects in the solar spectrum. A possible resolution of this time-scale problem was offered by the results of \citet{Baraffe10} who calculated pre-main-sequence models and found that if one assumed that the solar accretion occurred episodically the models could swiftly cross the Hayashi-track in the HR diagram and arrive at the main sequence with a shallow convection zone in less than a few Myr. Further calculations of accreting models by \citet{Vorobyov17}, \citet{Kunitomo17} and \citet{Kunitomo18} showed that the depth and time-scale of the early convection zone were highly dependent on the entropy and deuterium abundance of the accreted gas: if the gas retained most of its acquired heat from the gravitational collapse, and contained realistic amounts of deuterium to burn, the convection reached large depths, suggesting that the explanation for the solar \emph{ME} should be sought for elsewhere.    

\section{Stellar pairs}
\label{Pairs}

The advances in high-resolution spectroscopy in recent decades made it possible to systematically explore the dogma that stars, formed together in an interstellar cloud and now existing as components in binary stars or comoving pairs, should have very similar chemical composition, reflecting their common origin. Provided that the components have similar parameters (effective temperatures, surface gravities and over-all metallicities) accuracies on the order of 0.01 dex may be reached in the differential abundance estimates. Presently, about 19 such objects have been suggested to have stars showing significant mutual abundance differences, see Table~\ref{tab1}.

\begin{table}[h]
\caption{Stellar pairs with components with significantly different atmospheric abundances; the maximum logarithmic differences indicated by $\Delta_{\max}$}\label{tab1}%
\begin{tabular}{@{}lll@{}}
\toprule
Names & $\Delta_{\max}$  & Reference \\
\midrule
16 Cyg A, B  &  0.04 &  \citet{Ramirez11}, \citet{TucciMaia19} \\
XO-2         &  0.10 &  \citet{Biazzo15}, \citet{Ramirez15}  \\
WASP 94 A, B &  0.03 &  \citet{Teske16}  \\
HAT-P-4      &  0.1  &  \citet{Saffe17}  \\
HD 240430, HD 240429 &	0.2 & \citet{Oh18}, \citet{Miquelarena24} \\
HIP 71726, HIP 71737 &   0.1 & \citet{Galarza21} \\
6 wide binaries & 0.05--0.2 & \citet{Nagar20} \\    
7 co-moving pairs & 0.05--0.2 & \citet{Liu24}  \\    
\botrule
\end{tabular}
\end{table} 

\begin{figure}[ht]
\includegraphics[width=\textwidth]{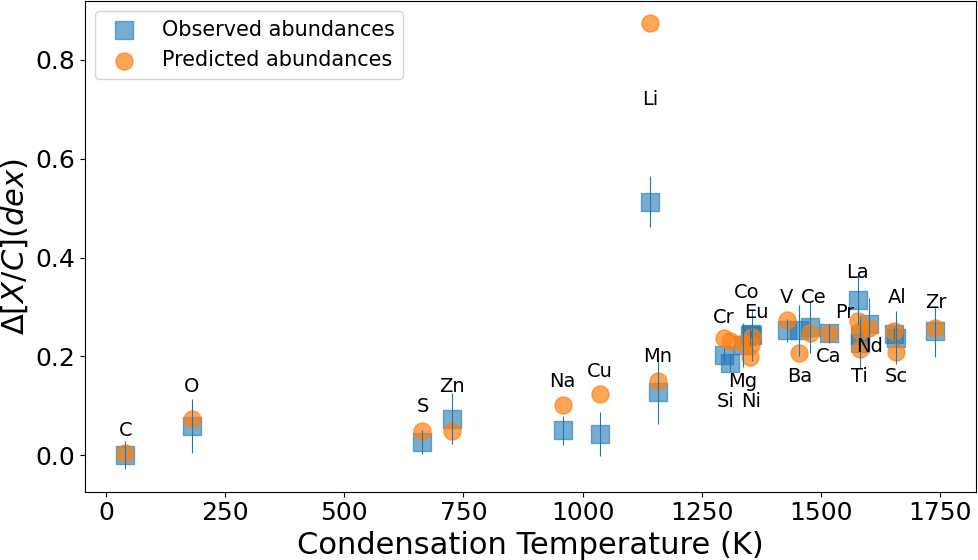}
\caption{The abundance differences between the co-moving stars HD\,240430 (Kronos) and HD\, 240429 (Krios) for elements plotted in order of increasing condensation temperature. Abundances are normalized on the carbon abundances. Observations are denoted by squares and predicted change in the surface abundance differences by oranges dots, assuming that rocky material with 19.9 Earth masses of the Earth bulk composition and 7.9 Earth masses of meteoritic composition, is added to Kronos and remains in its convection zone. Image reproduced with permission from \citet{Miquelarena24}, copyright by the author(s).}
\label{FigY2}
\end{figure}

For most of these stars, the differences $\Delta_e$ show a variation with the condensation temperature $T_c$ (for an example, see Fig. \ref{FigY2}). One might suggest that this phenomenon is related to the \emph{ME}, and that studies of it might elucidate the solar phenomenon. If so, an obvious indication is that the appearance of these differences in between two stars in the same system suggests that the \emph{ME} also is a local phenomenon, making the more global hypotheses discussed in Sects.~\ref{Evolution}--\ref{Turbulence} less probable. 

According to \citet{Liu24} the variation of $\Delta_e$ with the condensation temperature $T_c$ is found for 7 co-moving pairs out of a sample of 91 pairs. Among the 14 wide binaries explored by \citet{Nagar20}, 6 were found to have mutually different abundances and two of these with a significant $T_c$ dependence of $\Delta_e$. If this result is assumed to be representative also for single individual stars it suggests that about 4--7\% of all stars are affected by the \emph{ME} which is roughly consistent with the estimates given above in Sect.~\ref{Intro}.  

Just as for the \emph{ME} the differences in abundances of the binaries were soon related to planets, and specifically to the possibility that the component with more refractories had experienced an \textit{ingestion} of one or several rocky planets. Such events were proposed early by \citet{Sandquist98, Sandquist02}, for explaining high metallicities and unexpected Li abundances, but may also lead to considerable speed-ups of the stellar rotation rate (\citealt{Oetjens20}). A common idea is that the engulfment could be the result of perturbed orbits of the rocky planets, caused by inward migration by gas giants. In this scenario the Solar System rocky planets, like the Earth, was saved from engulfment by the locking of Jupiter and Saturn in the near-resonance of their mean motions. 

However, the hypothesis that the finding of ingestion signatures have also been criticized by \citet{Behmard23}, who argued that inhomogeneous stellar samples and analyses may have induced errors into the differential abundance  determinations.

These ingestion scenarios also raise time-scale problem: First, an ingested planet in a solar-type star should only leave markings in the photospheric spectrum from an excess of refractory elements for a limited time, since the excess of heavy elements will not remain in the convection zone.  They will be mixed into the radiative stellar interior by the (``fingering'') thermal haline instability at the bottom of the convection zone, a result of the gradient in mean-molecular weight counteracted by the entropy gradient, see \citet{Ulrich72}, \citet{Kippenhahn80}. This ``double diffusive instability'' develops since heat is exchanged more efficiently than molecular weight. The time scale of this mixing is about 1 Gyr \citep{Theado12, Behmard23} while the known binary stars and co-moving pairs with abundance differences have ages in the interval 2--8 Gyr.

These time scales would suggest that the rocky planets in the inner planetary systems must be able to remain in orbits for very considerable times before the ingestion episodes, in spite of the presence of giants in their neighborhood, if the migration of the giants occurs due to the friction in the proto-planetary disk. This is contrary to the results from simulations by Niemi \& Gustafsson (in preparation) suggesting that a migrating giant planet regularly passing smaller rocky planet within a distance less than 0.2 AU will link the latter into the host star, or scatter it out of the system within times typically smaller than $10^4$ years with a high probability. Alternatively, the ingestion could possibly be promoted long after the disintegration of the disk by internal instabilities in the planetary systems similar to those that occurred in the Solar System after about 1 Gyr, and led to the so-called Late Heavy Bombardment, or generated by instabilities in resonant chains of super-Earths and/or mini-Neptunes (\citealt{Izidoro17}).

In this connection one should note the interesting finding by \citet{Winter20} that Hot Jupiters (HJs) are much overrepresented among stars in co-moving star fields. Maybe, interactions between these stellar groups and giant molecular clouds drove the planetary migration via the Lidov-Kozai-Zeipel mechanism among the neighboring stars (Diederik Kruijssen \& Hans Rickman, private communication).The limited life-time of the co-moving star fields, with an upper limit set by \citet{Winter20} to 4.5 Gyr due to dynamical heating in the Milky Way, suggest that the high frequency of associations between the star fields and systems with HJs indicate that the life time of these planets are also limited, perhaps due to ingestion by their host stars.

Another time-scale is set by the limited time $t_w$ before the thermal-haline instability has erased the markings of planet engulfment in the stellar spectra, The frequency of binary components claimed to have excesses of refractories is ranging from 1 out of 35 \citep{Behmard23}, via 7 out of 91 \citep{Liu24} up to 33 of 76 \citep{Spina21}.  The age of the binary stars, and of the co-moving stars in the \citet{Liu24} sample extends from 2 to 8 Gyr. With a $t_w$ of 100 Myr to 1 Gyr and typical stellar life times of 5 Gyr it follows that more than 1/3 of all solar-type stars should have experienced this kind of event during their hydrogen core-burning phase on the Main Sequence. An interesting aspect of this is what records such an event in our Solar system could have left for us to explore.

The planet ingestion hypothesis for the origin of claimed abundance differences between the components in binaries or comoving pairs suggests a test, based on the fact that the mass of the hydrogen-ionisation convection zone deceases significantly with increasing effective temperature of the star. Therefore, one would expect systematically greater abundance shifts $\Delta_e$ when proceeding along the spectral sequence from G5 to F2. The sample of stars listed in Table 1, with its relatively small $T_{\rm eff}$ range from 5500 K to 6500 K, does not show any clear such tendency. It is also noteworthy that \citet{Ramirez14}, when exploring the variation of $\Delta_e$ with $T_{\rm eff}$ for single stars did not find any systematic difference in the tendency with $T_c$ when $T_{\rm eff}$ increased from 5950 K to 6300 K. An increase of the sample size for stellar pairs with a still wider $T_{\rm eff}$ range could be important; if the expected increase in $\Delta_e$ does not show up then, it lends support to other hypotheses such as the Nordlund selective outflow mechanism presented in Sect.~\ref{Outflows} which would affect the total chemical composition of the star and not only its surface layers. It should be noted, however, that \citet{Yong23} when analysing a complete census of co-moving stellar pairs with high-precision parameters found a significant tendency for the fraction of pairs with differences in [Fe/H] between the components to increase with effective temperature for $T_{\rm eff}>6000$\,K.

Also, as suggested by \citet{Soliman24}, dust-gas separation due to turbulence and radiation in the star-forming regions at small scales (down to 0.01 pc) might possibly explain the abundance differences between the components in the stellar pairs.

\FloatBarrier

The possibility to explain the \emph{ME} by advocating more frequent planet ingestions in solar twins than in the Sun, would require repeated linking orbits of rocky planets into their hosts  through many Gyrs, at least one planet per every Gyr for every star except for those belonging to the refractory-poor minority. Such a scenario seems improbable.

Recently, \citet{Yu25} \,\, have found that stars depleted in refractories in pairs with components deviating in chemical composition are magnetically more active than their companions. They tentatively ascribe this interesting correlation to star-planet interactions.

\section{Atomic diffusion and mixing in the stars}
\label{Diffusion}

The surface abundances in solar-type stars are known to be affected selectively by ``atomic diffusion'': gravitational settling when more heavy elements tend to sink through the gas while radiative acceleration may on the other hand levitate them towards the surface, depending on the detailed absorption spectrum of the atoms and ions of the element. However, turbulent mixing, rotational mixing as well as thermal-haline mixing also moderate these effects, see \citet{Michaud15}, to extents that are not yet predictable in full detail. 

Empirically, the effect of these phenomena show up in individual globular clusters as abundance differences between stars with different surface gravities and temperatures, as first demonstrated by \citet{Korn07}. The abundance differences between the stars mainly reflect the different mass of their convection regions. The effects have been modeled by \citet{Richard05} and \citet{Richard05a} with the effects on stellar evolution, and a grid of models for atmospheric abundance effects was published by \citet{Dotter17}. In those models the turbulent diffusion coefficient was scaled to fit the depletion in the metal-poor cluster NGC\,6397 observed by \citet{Korn07}. However, later studies of more metal-rich clusters suggest more efficient turbulent mixing for them, see in particular the detailed study of 86 stars in the cluster M\,4 ($[{\rm Fe/H}]= -1.13$) by \citet{Nordlander24}, which suggests that the abundance effects predicted by \citet{Dotter17} are exaggerated. The diffusion signatures found among the stars in rich open clusters (M\,67, \citealt{Liu19}, and NGC\,2420, \citealt{Semenova20}) with approximately solar metallicity suggest typically a reduction of the initial atmospheric iron abundance in the atmospheres of main-sequence stars at the turn-off point of about 0.1 dex. Thus, the effects are similar in magnitude to those of the \emph{ME} and one must ask whether the two effects can be separated.

As long as the comparison stars (``solar twins'') have fundamental parameters very close to those of the Sun one would not expect the effects of diffusion and mixing to interfere with the observed \emph{ME}, at least if the turbulent mixing is not an extra free parameter, for instance generated by differences in angular momentum distributions among and within the stars. (Note, however, that the requirement of similarity between the Sun and the stars also includes the age; the diffusion effect is time dependent.) However, in practice there are some differences in the positions of the Sun and the comparison stars in the parameter space where the abundance profiles caused by the different phenomena should be intercompared. 

\begin{figure}[ht]
\includegraphics[width=\textwidth]{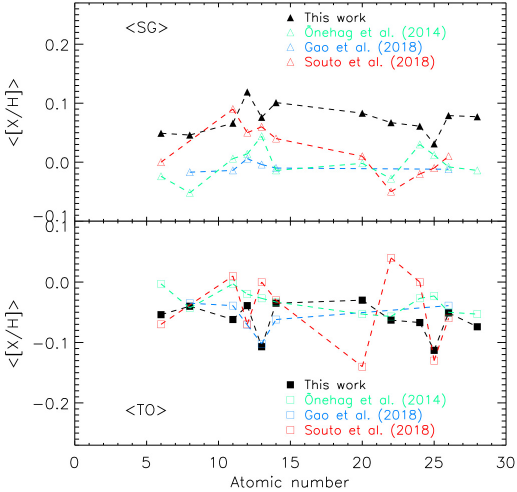}
\caption{Abundances of subgiants (upper panel) and Main Sequence stars at the Turn Off Point (lower panel) in the cluster M\,67, compared by \citet{Liu19} with results from three additional studies of the cluster included. The TOP stars tend to show lower abundances due to effects of atomic diffusion which are diluted for the sub giants by their deeper convection zone. Image reproduced with permission from \citet{Liu19}, copyright by ESO.}
\label{FigY3}
\end{figure}

The predicted profiles of depletion for solar-type dwarfs of \citet{Dotter17} with radiative acceleration and turbulence included suggest depletion effects  of about -0.1 dex in the iron abundance, and smaller effects for O, Mg, Si and Ca. The effects for NGC\,2420 observed by \citet{Semenova20} are about $-0.1$ to -0.2 dex for Fe, Ca and Mg with a considerable scatter. \citet{Liu19} traced effects of similar magnitudes for the turn-off stars of M\,67, for about 20 different elements, including volatiles like C, O, Na, S, and Zn, and a number of refractories including Al (see Fig. \ref{FigY3}). \citet{Bertelli18} also found similar signatures in main-sequence solar-type stars in M 67 for the elements C, Na, Mg, Al, Si, Ca TI, Cr, Mn and Fe. For stars with a surface gravity of log $g \approx 4.2$ they obtained abundance ratios [X/Fe] reduced by 0.1 dex relative to stars with solar gravity (about 4.5), and similar reductions for both elements X = Al and Mn, a pronounced refractory and a volatile, respectively. For the volatiles C and O, the effects on [X/Fe] are as a mean close to 0.0 dex. The hypothesis that a non-solar and positive logarithmic ratio of refractories to volatiles (the \emph{ME}) would be caused by diffusion is not supported by these findings. 

Calculations by Olivier Richard (private communication) with gravitational settling and radiative levitation suggest a decrease of the surface abundances relative to hydrogen for stars with solar mass evolving on the main sequence of less than 0.1 dex. The variations in between the different elements are small, with the exception of potassium (condensation temperature of 1001\,K) and chromium (1291\,K) which show variations of the order of 0.02 dex. The effects are half of this if turbulent mixing is added to reproduce the decrease of the lithium abundance with time. The relative abundance effects on volatiles versus refractories are only about 0.01 dex or less in these simulations.

A given star may, however, be affected by both diffusion/mixing and the \emph{ME}, and to disentangle which is the dominating reason could be intricate if the effects are small. In their detailed study of seven binary stars \citet{Liu21} concluded by comparing abundances for the different components of the pairs, that four of the pairs showed subtle abundance differences of 0.01 -- 0.03 dex. The authors tentatively ascribed these differences to diffusion. Among those four pairs were two that host known planets. The remaining three pairs showed more significant abundance differences with clear trends with $T_c$ which can then be attributed to the \emph{ME}.        

\FloatBarrier     

\section{Relations to life on Earth?}
\label{Life} 
 
The original application from 2007 by Mel{\'e}ndez et al. for observing time for spectra which led to the discovery of the \emph{ME} had the title \textit{The Fundamental Blocks of Life in Solar Twins and Planet-hosting Stars}. In the \emph{Scientific objectives} the Observing Programmes Committee could read: ``The biogenic elements H, C, N, O, S, and P are the essential ingredients for life as we know it. We propose to undertake the first observational survey of biogenic elements in solar twins and in stars''. 
Certainly, not all members of our research group, and probably nor of the OPC, agreed that such astrobiological rationale were the strongest aspects of the application. Anyhow, it is interesting to note that the result of the study, as seen in Fig.~\ref{Fig1}, shows an excess of the biogenic elements for the Sun as compared with the twins. 

How, if at all, does our result relate to the existence of life on our planet? Could one believe in an ``anthropic'' reason for the \emph{ME}, postulate that there is a general spread in abundances of biogenic elements relative to refractories among the solar-type stars and that life -- and humans -- were formed on Earth \textit{because} the pre-solar elemental composition was most benign for us?  If so, an issue is whether the relatively small excesses by at the most 20\% of the biogenic elements, as compared with the more common twins, really increase the possibilities for life. This would require some kind of thresholds in abundance ratios that have to be passed for life to form. Contacts with specialists in the early chemical evolution of life have suggested that there are no known definitive thresholds in abundances of these elements as long as the PH value is kept in the interval 6--8, which may well be more an issue of geochemistry than precise initial chemical composition (\citealt{Krishnamurthy18} and private communication). Also, one might question whether really S and P, as well as K and Na, are at all absolutely necessary for life-like structures to form or whether these elements could be substituted by others (\citealt{Goldford17}, \citealt{Athavale12}). 

Other conditions for life on Earth, which might be related to the chemical abundances, are geophysical such as the dependences on the geomagnetic field. That field, important to protect life from unhealthy cosmic radiation, is probably dependent on the iron abundance. The field, with its origin in the iron core of the Earth, would perhaps gain strength from an increased iron abundance which is contrary to the finding of the \emph{ME}: a comparatively low refractory abundance of the Sun.  We note, however, ``the nucleation paradox'' of the core \citep{Huguet18}: it is found that the inner core is solid, surrounded by a iron liquid with convection motions driven by heat from the inner solidification and driving the dynamo of the magnetic field. This state of the core is enigmatic: one would expect the core to be supercooled and fully fluent. Perhaps inhomogeneities or pollutions of oxides from the mantle could explain the state \citep{Wilson24}. Whether this in some way is related in to the detailed chemical composition of the matter from which the Earth originated is highly questionable. 

Another more astrophysical circumstance of potential significance for life on Earth has already been touched upon: the mean-motion resonances of Jupiter and Saturn. When, in the \textit{Grand Tack} model \citep{Raymond14} Jupiter was migrating inwards toward the Earth, but then was “picked up” by Saturn and moved out again, thanks to 3:2, 2:1 and a following near 5:2 resonance, the inner rocky planets of our Solar System were left in peace, and still are.  Also, Jupiter early on may have hindered super-earths to move inwards \citep{Izidoro15}. All this may have been fortunate and perhaps critically dependent on the initial conditions in the proto-planetary disk. The excess of biogenic elements did probably not enhance the chances for observers to form (and raise the issue of the 
\emph{ME}), but may, as seen above, be caused by the location of the gas giants which was also a probable condition for life to evolve on the Earth. So, there could anyhow be a common indirect link between the \emph{ME} and our existence. 

\section{Conclusions}
\label{Conc}
So, what are the reasons for the \emph{ME}? No definitive answer has been possible to provide, as yet. Galactic evolution and solar migration cannot be excluded (Sect.~\ref{Evolution}). Further studies relating ages,space motions and detailed abundance measurements for stars in the inner Galaxy as well as in the solar neighborhood will be needed to clarify the situation. Obviously, more observations are needed, especially of stars in well controlled samples to ascertain that the Sun is really compared with contemporary “twins”.

The radiative cleansing of the proto-solar nebula at different stages of the star forming process (Sect.~\ref{SF region} and Sect.~\ref{Self-cleansing}) may be a possibility but are not a very probable explanation. Attempts to explore the thin dust-cleansed but yet neutral region around H II regions, and possible star-formation therein, may anyhow be worthwhile, as well as searching for late infall of gas towards proto-stars and to study the possibilities for element fractionation in such flows The determination of accurate abundances of solar-like stars in old and rich Population I clusters like M\,67 and NGC\,188 should proceed. The separation of dust and gas by turbulence in molecular clouds (Sect.~\ref{Turbulence}) still offers possible explanations for the \emph{ME}, in particular when gravitation, radiation and magnetic fields are considered. Models and detailed high-resolution observations of such clouds could clarify this. 

Separation of gas and dust in protoplanetary disks (Sects. \ref{Separation} and \ref{Outflows}) may occur in several ways. No doubt these mechanisms offer promising possibilities to explain the \emph{ME} even though more specific observable signatures of these phenomena would be important. Some of these mechanisms may be little dependent on the planets \emph{per se} but governed by the different strong outflows from the disks which may lead to global abundance differences between the stars, including their interiors. Also the forming planets may be of significance: the special role of the gas giants when kept outside the innermost planetary system and outside the zone where dust and pebbles would evaporate should be further explored (Sect.~\ref{Gas giants}), while also the formation of rocky planets might possibly be significant as deposits of refractories (Sect.~\ref{Rocky planets}).  

The study of stellar pairs with common origins but different composition may be important for resolution of the \emph{ME} enigma (Sect.~\ref{Pairs}). However, the phenomena behind those binaries are not necessarily identical with those that led to the solar effect, and nor are they necessarily easier to understand. It is promising, though, that an extensive and rapidly growing sample of pairs offers interesting possibilities.

Gravitational settling, radiative levitation and mixing are now well established mechanisms and known to play important roles in the upper layers of solar-type stars. Yet, it is not very probable that these phenomena contribute importantly to the \emph{ME} with the possible exception of rotationally induced mixing, where the different angular momenta among the solar twins might induce abundance differences.   

The \emph{ME} and the existence of life on our planet are probably not related directly to the particular abundances but may sooner be indirectly linked via the coupling of Jupiter and Saturn in orbits far away from the orbit of the Earth in the Solar System (Sect.~\ref{Life}). However, the discovery of the \emph{ME} and the pursued interest with hundreds of studies in the last 15 years has illustrated how a speculative connection between the effect and the existence of habitable planets have led to a host of important new results and interesting new problems. This may in itself be regarded a good motivation for astrobiology and even SETI. Wild speculations should not be generally despised: they may serve as drivers of unexpected interesting discoveries and new solid science.

As is seen from the list above, there are many possible explanations for the \emph{ME} and also a number of maybe less attractive ones, although several of those are still not impossible. Indeed, some of these phenomena may even be at play simultaneously. We have not succeeded in following the stepwise elimination scheme of Sherlock Holmes, presented in the introductory quote. This is not an uncommon situation in astrophysics since we usually deal with complex systems. Several different approaches towards a solution of the problem need to be pursued in view of the fact that we do not know yet what interesting astrophysics would result. This situation illustrates not only a philosophical problem in contemporary science, but also a practical one: High-quality data leads to 
more complex structures being uncovered, which in turn requires more complicated models to provide understanding. Simple and striking explanations must then often be replaced by a multitude of intricate details.\footnote{One could note that the often quoted saying ``Elementary, my dear Watson!'', was never written by Conan Doyle, but seems to have been first introduced by P.G. Wodehouse in a book from 1915 \citep{Redmond94}.} It takes a great amount of patience to elucidate and clarify these phenomena,
and assured intuition to bring study to a halt at a relevant point.

\bmhead{Acknowledgements}
The present article originated as a review given at the conference \textit{Are we a unique species on a unique planet? -- Or just the ordinary Galactic standard?}  in Copenhagen July 30-August 2, 2024. Uffe Graae J{\o}rgensen, Anja C. Andersen and the other organizers are warmly thanked for arranging the meeting and inviting my talk. Anish Amarsi, Bengt Edvardsson, Aaron Goldman, Anders Johansen, Andrea Korn, Ramanarayanan Krishnamurty, Fan Liu, Jorge Mel{\'e}ndez, Toivo T. Niemi, {\AA}ke Nordlund, Olivier Richard and Hans Rickman, are thanked for comments, valuable discussions and suggestions. The referees are thanked for many helpful comments which considerably improved the article. Economical travel support was contributed by the Royal Science Society in Uppsala. 

\phantomsection
\addcontentsline{toc}{section}{References}

\bibliography{SunCny}

\end{document}